\pdfoutput=1

\documentclass[prd,
               twocolumn,
               amssymb,
               amsmath,
               eqsecnum,
               showpacs,
               letterpaper,
               superscriptaddress,
               altaffilletter]
               {revtex4}

\usepackage{color}
\usepackage{graphicx}
\usepackage{latexsym}
\usepackage{float}
\usepackage{amsmath}
\usepackage{amssymb}
\usepackage{multirow}

\renewcommand{\today}{\number\day\space\ifcase\month\or
  January\or February\or March\or April\or May\or June\or
  July\or August\or September\or October\or November\or December\fi
  \space\number\year}

\def\be{\begin{equation}}
\def\ee{\end{equation}}
\def\bi{\begin{itemize}} 
\def\ei{\end{itemize}}
\def\ben{\begin{enumerate}}
\def\een{\end{enumerate}}

\def\Zcoh{Z_{\text{H}_+}}
\def\Zcor{Z^{corr}_{\text{H}_+}}
\def\sfyo{S5y1}
\def\sfyt{S5y2}
\def\ligodoc{{LIGO-P09}{00108}-{v6}} 

\begin{document}

\title{All-sky search for gravitational-wave bursts in the first joint LIGO-GEO-Virgo run}

\begin{abstract}
\vspace*{0.2in}
We present results from an all-sky search for unmodeled gravitational-wave bursts
in the data collected by the LIGO, GEO\,600 and Virgo detectors between
November 2006 and October 2007. The search is performed by three
different analysis algorithms over the frequency band 50 -- 6000\,Hz. 
Data are analyzed for times with at least two of the four LIGO-Virgo detectors
in coincident operation, with a total live time of 266 days.
No events produced by the search algorithms survive the selection cuts.
We set a frequentist upper limit on the rate of gravitational-wave bursts
impinging on our network of detectors. When combined with the previous
LIGO search of the data collected  between November 2005 and 
November 2006,
the upper limit on the rate of detectable gravitational-wave
bursts in the 64--2048 Hz band
is 2.0 events per year at 90\% confidence. We also present
event rate versus strength exclusion plots for several types of plausible
burst waveforms.
The sensitivity of the combined search is expressed in terms of 
the root-sum-squared strain amplitude
for a variety of simulated waveforms and lies in the range
$6 \times 10^{-22}$ Hz$^{-1/2}$ to $2 \times 10^{-20}$ Hz$^{-1/2}$. 
This is the first untriggered burst search to use data from the LIGO and
Virgo detectors together, and the most sensitive untriggered burst search
performed so far.
\end{abstract}

\affiliation{Albert-Einstein-Institut, Max-Planck-Institut f\"ur Gravitationsphysik, D-14476 Golm, Germany$^\ast$}
\affiliation{Albert-Einstein-Institut, Max-Planck-Institut f\"ur Gravitationsphysik, D-30167 Hannover, Germany$^\ast$}
\affiliation{Andrews University, Berrien Springs, MI 49104 USA$^\ast$}
\affiliation{AstroParticule et Cosmologie (APC), CNRS: UMR7164-IN2P3-Observatoire de Paris-Universit\'e Denis Diderot-Paris 7 - CEA : DSM/IRFU$^\dagger$}
\affiliation{Australian National University, Canberra, 0200, Australia$^\ast$}
\affiliation{California Institute of Technology, Pasadena, CA  91125, USA$^\ast$}
\affiliation{Caltech-CaRT, Pasadena, CA  91125, USA$^\ast$}
\affiliation{Cardiff University, Cardiff, CF24 3AA, United Kingdom$^\ast$}
\affiliation{Carleton College, Northfield, MN  55057, USA$^\ast$}
\affiliation{Charles Sturt University, Wagga Wagga, NSW 2678, Australia$^\ast$}
\affiliation{Columbia University, New York, NY  10027, USA$^\ast$}
\affiliation{European Gravitational Observatory (EGO), I-56021 Cascina (Pi), Italy$^\dagger$}
\affiliation{Embry-Riddle Aeronautical University, Prescott, AZ   86301 USA$^\ast$}
\affiliation{E\"otv\"os University, ELTE 1053 Budapest, Hungary$^\ast$}
\affiliation{ESPCI, CNRS,  F-75005 Paris, France$^\dagger$}
\affiliation{Hobart and William Smith Colleges, Geneva, NY  14456, USA$^\ast$}
\affiliation{$^a$INFN, Sezione di Firenze, I-50019 Sesto Fiorentino, Italy$^\dagger$\\$^{17b}$Universit\`a degli Studi di Urbino 'Carlo Bo', I-61029 Urbino, Italy$^\dagger$}
\affiliation{INFN, Sezione di Genova;  I-16146  Genova, Italy$^\dagger$}
\affiliation{$^a$INFN, sezione di Napoli, I-80126 Napoli, Italy$^\dagger$\\$^{19b}$Universit\`a di Napoli 'Federico II' Complesso Universitario di Monte S.Angelo, I-80126 Napoli, Italy$^\dagger$\\$^{19c}$Universit\`a di Salerno, Fisciano, I-84084 Salerno, Italy$^\dagger$}
\affiliation{$^a$INFN, Sezione di Perugia, I-6123 Perugia, Italy$^\dagger$\\$^{20b}$Universit\`a di Perugia, I-6123 Perugia, Italy$^\dagger$}
\affiliation{$^a$INFN, Sezione di Pisa, I-56127 Pisa, Italy$^\dagger$\\$^{21b}$Universit\`a di Pisa, I-56127 Pisa, Italy$^\dagger$\\$^{21c}$Universit\`a di Siena, I-53100 Siena, Italy$^\dagger$}
\affiliation{$^a$INFN, Sezione di Roma, I-00185 Roma, Italy$^\dagger$\\$^{22b}$Universit\`a 'La Sapienza', I-00185  Roma, Italy$^\dagger$}
\affiliation{$^a$INFN, Sezione di Roma Tor Vergata, Italy$^\dagger$\\$^{23b}$Universit\`a di Roma Tor Vergata, Italy$^\dagger$\\$^{23c}$Universit\`a dell'Aquila, I-67100 L'Aquila, Italy$^\dagger$}
\affiliation{Institute of Applied Physics, Nizhny Novgorod, 603950, Russia$^\ast$}
\affiliation{Inter-University Centre for Astronomy and Astrophysics, Pune - 411007, India$^\ast$}
\affiliation{LAL, Universit\'e Paris-Sud, IN2P3/CNRS, F-91898 Orsay, France$^\dagger$}
\affiliation{Laboratoire d'Annecy-le-Vieux de Physique des Particules (LAPP),  IN2P3/CNRS, Universit\'e de Savoie, F-74941 Annecy-le-Vieux, France$^\dagger$}
\affiliation{Leibniz Universit\"at Hannover, D-30167 Hannover, Germany$^\ast$}
\affiliation{LIGO - California Institute of Technology, Pasadena, CA  91125, USA$^\ast$}
\affiliation{LIGO - Hanford Observatory, Richland, WA  99352, USA$^\ast$}
\affiliation{LIGO - Livingston Observatory, Livingston, LA  70754, USA$^\ast$}
\affiliation{LIGO - Massachusetts Institute of Technology, Cambridge, MA 02139, USA$^\ast$}
\affiliation{Laboratoire des Mat\'eriaux Avanc\'es (LMA), IN2P3/CNRS, F-69622 Villeurbanne, Lyon, France$^\dagger$}
\affiliation{Louisiana State University, Baton Rouge, LA  70803, USA$^\ast$}
\affiliation{Louisiana Tech University, Ruston, LA  71272, USA$^\ast$}
\affiliation{McNeese State University, Lake Charles, LA 70609 USA$^\ast$}
\affiliation{Montana State University, Bozeman, MT 59717, USA$^\ast$}
\affiliation{Moscow State University, Moscow, 119992, Russia$^\ast$}
\affiliation{NASA/Goddard Space Flight Center, Greenbelt, MD  20771, USA$^\ast$}
\affiliation{National Astronomical Observatory of Japan, Tokyo  181-8588, Japan$^\ast$}
\affiliation{$^a$Nikhef, National Institute for Subatomic Physics, P.O. Box 41882, 1009 DB Amsterdam, The Netherlands$^\dagger$\\$^{41b}$VU University Amsterdam, De Boelelaan 1081, 1081 HV Amsterdam, The Netherlands$^\dagger$}
\affiliation{Northwestern University, Evanston, IL  60208, USA$^\ast$}
\affiliation{$^a$Universit\'e Nice-Sophia-Antipolis, CNRS, Observatoire de la C\^ote d'Azur, F-06304 Nice, France$^\dagger$\\$^{43b}$Institut de Physique de Rennes, CNRS, Universit\'e de Rennes 1, 35042 Rennes, France$^\dagger$}
\affiliation{$^a$INFN, Gruppo Collegato di Trento, Trento, Italy$^\dagger$\\$^{44b}$Universit\`a di Trento,  I-38050 Povo, Trento, Italy$^\dagger$\\$^{44c}$INFN, Sezione di Padova, I-35131 Padova, Italy$^\dagger$\\$^{44d}$Universit\`a di Padova, I-35131 Padova, Italy$^\dagger$}
\affiliation{$^a$IM-PAN, 00-956 Warsaw, Poland$^\dagger$\\$^{45b}$Warsaw University, 00-681 Warsaw, Poland$^\dagger$\\$^{45c}$Astronomical Observatory of Warsaw University, 00-478 Warsaw, Poland$^\dagger$\\$^{45d}$CAMK-PAN, 00-716 Warsaw, Poland$^\dagger$\\$^{45e}$Bia{\l}ystok University, 15-424 Bia{\l}ystok, Poland$^\dagger$\\$^{45f}$IPJ, 05-400 \'Swierk-Otwock, Poland$^\dagger$\\$^{45g}$Institute of Astronomy, 65-265 Zielona G\'ora, Poland$^\dagger$}
\affiliation{Rochester Institute of Technology, Rochester, NY  14623, USA$^\ast$}
\affiliation{Rutherford Appleton Laboratory, HSIC, Chilton, Didcot, Oxon OX11 0QX United Kingdom$^\ast$}
\affiliation{San Jose State University, San Jose, CA 95192, USA$^\ast$}
\affiliation{Sonoma State University, Rohnert Park, CA 94928, USA$^\ast$}
\affiliation{Southeastern Louisiana University, Hammond, LA  70402, USA$^\ast$}
\affiliation{Southern University and A\&M College, Baton Rouge, LA  70813, USA$^\ast$}
\affiliation{Stanford University, Stanford, CA  94305, USA$^\ast$}
\affiliation{Syracuse University, Syracuse, NY  13244, USA$^\ast$}
\affiliation{The Pennsylvania State University, University Park, PA  16802, USA$^\ast$}
\affiliation{The University of Melbourne, Parkville VIC 3010, Australia$^\ast$}
\affiliation{The University of Mississippi, University, MS 38677, USA$^\ast$}
\affiliation{The University of Sheffield, Sheffield S10 2TN, United Kingdom$^\ast$}
\affiliation{The University of Texas at Austin, Austin, TX 78712, USA$^\ast$}
\affiliation{The University of Texas at Brownsville and Texas Southmost College, Brownsville, TX  78520, USA$^\ast$}
\affiliation{Trinity University, San Antonio, TX  78212, USA$^\ast$}
\affiliation{Universitat de les Illes Balears, E-07122 Palma de Mallorca, Spain$^\ast$}
\affiliation{University of Adelaide, Adelaide, SA 5005, Australia$^\ast$}
\affiliation{University of Birmingham, Birmingham, B15 2TT, United Kingdom$^\ast$}
\affiliation{University of Florida, Gainesville, FL  32611, USA$^\ast$}
\affiliation{University of Glasgow, Glasgow, G12 8QQ, United Kingdom$^\ast$}
\affiliation{University of Maryland, College Park, MD 20742 USA$^\ast$}
\affiliation{University of Massachusetts - Amherst, Amherst, MA 01003, USA$^\ast$}
\affiliation{University of Michigan, Ann Arbor, MI  48109, USA$^\ast$}
\affiliation{University of Minnesota, Minneapolis, MN 55455, USA$^\ast$}
\affiliation{University of Oregon, Eugene, OR  97403, USA$^\ast$}
\affiliation{University of Rochester, Rochester, NY  14627, USA$^\ast$}
\affiliation{University of Salerno, 84084 Fisciano (Salerno), Italy$^\ast$}
\affiliation{University of Sannio at Benevento, I-82100 Benevento, Italy$^\ast$}
\affiliation{University of Southampton, Southampton, SO17 1BJ, United Kingdom$^\ast$}
\affiliation{University of Strathclyde, Glasgow, G1 1XQ, United Kingdom$^\ast$}
\affiliation{University of Western Australia, Crawley, WA 6009, Australia$^\ast$}
\affiliation{University of Wisconsin--Milwaukee, Milwaukee, WI  53201, USA$^\ast$}
\affiliation{Washington State University, Pullman, WA 99164, USA$^\ast$}
\author{J.~Abadie$^\text{29}$}\noaffiliation\author{B.~P.~Abbott$^\text{29}$}\noaffiliation\author{R.~Abbott$^\text{29}$}\noaffiliation\author{T.~Accadia$^\text{27}$}\noaffiliation\author{F.~Acernese$^\text{19a,19c}$}\noaffiliation\author{R.~Adhikari$^\text{29}$}\noaffiliation\author{P.~Ajith$^\text{29}$}\noaffiliation\author{B.~Allen$^\text{2,77}$}\noaffiliation\author{G.~Allen$^\text{52}$}\noaffiliation\author{E.~Amador~Ceron$^\text{77}$}\noaffiliation
% ...

\collaboration{$^\ast$The LIGO Scientific Collaboration and $^\dagger$The Virgo Collaboration}
\noaffiliation

\date[\relax]{Dated: \today }

\maketitle

\section{Introduction}\label{sec:introduction}

The LIGO Scientific Collaboration (LSC) and the Virgo Collaboration
operate a network of interferometric gravitational-wave (GW) detectors
with the goal of detecting
gravitational waves from astrophysical sources.
Some of these sources may produce transient ``bursts'' of GW radiation
with relatively short duration ($\lesssim$1 s).
Plausible burst sources~\cite{cutler02} include merging compact binary
systems consisting of black holes and/or neutron stars~\cite{bbh, bhns},
core-collapse supernovae~\cite{ott2008}, neutron star 
collapse~\cite{SNwave}, starquakes associated with magnetar
flares~\cite{SGR} or pulsar glitches~\cite{glitches}, cosmic string
cusps~\cite{cusp}, and other violent events in the Universe.

During the most recent data-taking run five GW detectors were 
operational.  The three LIGO detectors~\cite{S5} started their  
Science Run 5 (S5) in November 2005, and the GEO\,600 detector~\cite{Grote} 
joined the S5 run in January 2006.  The Virgo detector~\cite{acernese2006} 
began its Virgo Science Run 1 (VSR1) in May 2007.  All five instruments took 
data together until the beginning of October 2007.

An all-sky search for GW burst signals has already been conducted on 
the first calendar year of the LIGO S5 data (referred to as ``\sfyo'')
in a wide frequency band of $64-6000$ Hz \cite{S5y1Burst, S5y1BurstHF}.
In this paper, we report on a search for GW burst signals in the frequency band
$50-6000$~Hz for the rest of the S5/VSR1 run, referred to as ``\sfyt/VSR1''. 
It includes data collected by the LIGO and Virgo detectors, 
which had comparable sensitivities, and uses three different search algorithms. 
In comparison with the \sfyo~analysis, the network of LIGO and Virgo detectors,  
spread over three sites, provides better sky 
coverage as well as improved capabilities to reject spurious signals. 
\sfyt/VSR1 is also the first long-term observation with 
the world-wide network of interferometric detectors. 
This is a major step forward with respect to previous 
observations led by the network of resonant detectors~\cite{astone2003, IGEC2-2007}, 
since, as we will show in this paper, the performance is improved 
by more than one order of magnitude both in the analyzed frequency
 bandwidth and the level of instrumental noise.

This paper is organized as follows.  
In Section~\ref{sec:s5detectors} we describe the LSC and Virgo instruments.  
In Section~\ref{sec:overview} we give a brief overview of the search procedure.   
In Section~\ref{sec:pipeline} we present the search algorithms. 
Simulations are described in
Section~\ref{sec:simulations}, and the error analysis in
Section~\ref{sec:systematics}. The results of the search are
presented in Section~\ref{sec:results}, and astrophysical implications are discussed in Section~\ref{sec:summaries}. The appendices provide additional details
on data characterization and the analysis pipelines.

\section{Detectors}\label{sec:s5detectors}

\subsection{LIGO}

LIGO consists of three detectors at two observatories in the
United States.  Each detector is a large Michelson-type
interferometer with additional mirrors forming Fabry-Perot cavities in
the arms and a power-recycling mirror in the input beam path.
Interferometric sensing and feedback is used to ``lock'' the
mirror positions and orientations to keep all of the optical
cavities on resonance.
A gravitational wave is sensed as a quadrupolar strain, measured
interferometrically as an effective difference between the lengths of
the two arms.
The LIGO Hanford Observatory, in Washington, houses
independent detectors with the arm lengths of 4 km and 2 km, called H1 and
H2 respectively.  The LIGO Livingston Observatory, in Louisiana, has
a single detector with 4-km arms, called L1.  The detector
instrumentation and operation are described in detail
elsewhere~\cite{S5}, and the improvements leading up to the S5 run
which are most relevant for GW burst searches have been described in
the first-year search~\cite{S5y1Burst}.

The best achieved sensitivities of the LIGO detectors during the second year of
S5, as a function of signal frequency, are shown in
Fig.~\ref{fig:Shh}.  The detectors are most sensitive over a band
extending from about 40\,Hz to a few kHz.  Seismic noise dominates at
lower frequencies since the effectiveness of the seismic isolation
system is a very strong function of frequency.  Above $\sim$200\,Hz,
laser shot noise corrected for the Fabry-Perot cavity response yields
an effective strain noise that rises linearly with frequency.  The sensitivity at
intermediate frequencies is determined mainly by thermal noise, with
contributions from other sources.  The peaks at $\sim$350\,Hz and
harmonics are the thermally-excited vibrational modes of the wires
from which the large mirrors are suspended.  Smaller peaks are due to
other mechanical resonances, power line harmonics, and calibration
signals.

Commissioning periods during the
second year of S5 led to incremental improvements in the detector
sensitivities.  The most significant of these were in January 2007,
when the seismic isolation systems at both sites were improved to
reduce the coupling of microseismic noise to the mirror suspensions,
thereby mitigating noise from the nonlinear Barkhausen
effect~\cite{BarkhausenExpts} in the magnets used to control the
mirror positions; and in August 2007, when the L1 frequency
stabilization servo was re-tuned.
Overall,
the average sensitivities of the H1 and L1 detectors during the second
year were about 20\% better than the first-year averages, while the
H2 detector (less sensitive to begin with by a factor of $\sim$2) had
about the same average sensitivity in both years.
The operational duty cycles for all three detectors also improved
as the run progressed, from (72.8\%, 76.7\%, 61.0\%) averaged over
the first year to (84.0\%, 80.6\%, 73.6\%) averaged over the second
year for H1, H2, and L1, respectively.

\subsection{GEO\,600}

The GEO\,600 detector, located near Hannover, Germany, also operated 
during S5, though with a lower sensitivity than the LIGO and Virgo 
detectors. The GEO\,600 data are not used in the initial search stage of
the current study as the modest gains in the sensitivity to GW signals
would not offset the increased complexity of the analysis. The GEO\,600 data
are held in reserve, and used to follow up any detection candidates from
the LIGO-Virgo analysis.

GEO\,600 began its participation in S5 on January~21 2006, acquiring 
data during nights and weekends.  Commissioning work was performed 
during the daytime, focussing on gaining a better
understanding of the detector and improving data quality.
GEO switched to full-time data taking from May~1 to October~6, 2006, 
then returned to night-and-weekend mode through the end of the S5 run.
Overall GEO\,600 collected about 415 days of science data during S5, 
for a duty cycle of 59.7\% over the full S5 run.

\subsection{Virgo}

The Virgo detector~\cite{acernese2006}, also called V1, is an 
interferometer with 3 km arms located near Pisa in Italy.
One of the main instrumental differences with respect to LIGO is 
the seismic isolation system based on super-attenuators~\cite{superattenuators}, 
chains of passive attenuators capable of filtering seismic disturbances in 6
degrees of freedom with sub-Hertz corner frequencies.
For VSR1, 
the Virgo duty cycle was 81\% and the longest
continuous period with the mirror positions interferometrically controlled
was more than $94$ hours.
Another benefit from super-attenuators is a significant reduction of the detector noise
at very low frequency ($< 40$~Hz) where Virgo surpasses the LIGO sensitivity. 

\begin{figure}
\begin{center}
\mbox{
\includegraphics*[width=0.5\textwidth]{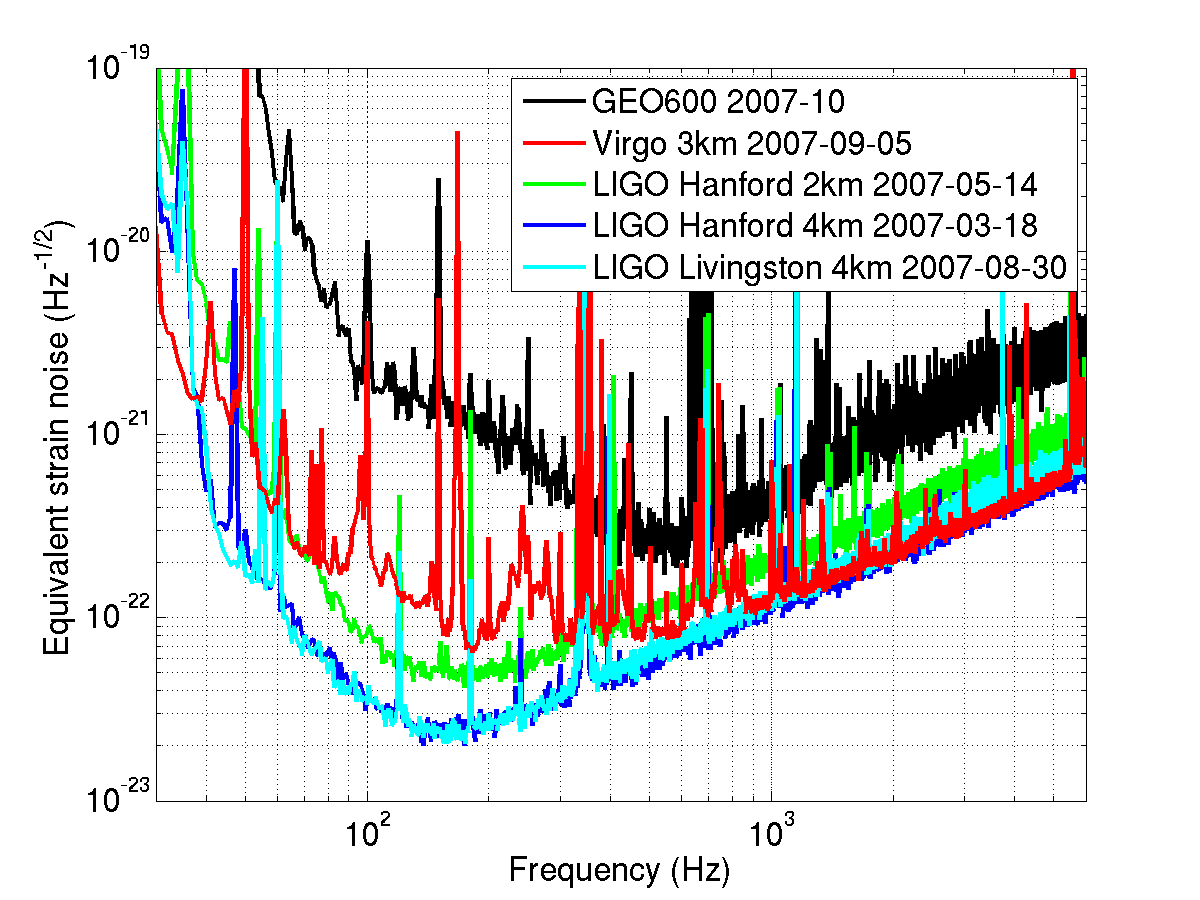}}
\caption{Best noise amplitude spectral densities of the five
LSC/Virgo detectors during S5/VSR1.}
\label{fig:Shh}
\end{center}
\end{figure}

Above $300$~Hz, the spectral sensitivity achieved by Virgo during VSR1 
is comparable to that of LIGO (see Figure~\ref{fig:Shh}). Above $500$~Hz 
the Virgo sensitivity is dominated by shot noise.
Below $500$~Hz there is excess noise 
due to environmental and instrumental noise sources, and below  
300~Hz these produce burst-like transients. 

Due to the different orientation of its arms, the antenna pattern 
(angular sensitivity) of Virgo is complementary to that of the LIGO 
detectors, with highest response in directions of low LIGO sensitivity.
Virgo therefore significantly increases
the sky coverage of the network. In addition, simultaneous observations 
with the three LIGO-Virgo
sites improve rejection of spurious signals and allow reconstruction of the sky position 
and waveforms of detected GW sources. 

\section{Search Overview}\label{sec:overview}

The analysis described in this paper uses data from the
LIGO detectors collected from 14 November 2006 
through 1 October 2007 (\sfyt), 
and Virgo data from VSR1,   
which started on 18 May 18 2007 
and ended at the same time as S5 \footnote{This search includes several
hours of data collected on 1 October 2007, after the official end of the S5
and VSR1 runs at 0:00 UTC on that day.}.
The procedure used for this \sfyt/VSR1 search is the same as that used 
for \sfyo~\cite{S5y1Burst}.  In this section we briefly review the main 
stages of the analysis.

\subsection{Data quality flags}
\label{sec:dq}

The detectors are occasionally affected by
instrumental or data acquisition artifacts as well as by periods of
degraded sensitivity or an excessive rate of transient noise due 
to environmental conditions such as bad weather.  
Low-quality data segments are tagged with 
Data Quality Flags (DQFs). 
These DQFs are divided into three categories depending on
their seriousness.  Category 1 DQFs are used to define the data segments processed 
by the analysis algorithms.  Category 2 DQFs are unconditional data cuts 
applied to any events generated by the algorithms.  Category 3 DQFs define the 
clean data set used to calculate upper limits on the GW rates.

We define DQFs for \sfyt/VSR1 following the approach used for \sfyo~\cite{S5y1Burst}.  
More details are given in Appendix~\ref{sec:dataquality}.
After category 2 DQFs have been applied,
the total available time during this period
is 261.6 days for H1, 253.4 days for H2, 233.7 days for L1 and
106.2 days for V1~\footnote{For reference, a list of all time intervals
included in the search may be found at https://dcc.ligo.org/cgi-bin/DocDB/ShowDocument?docid=T1000099\&version=1 .}.

\subsection{Candidate Event Generation}

As discussed in Section~\ref{sec:pipeline}, three independent 
search algorithms are used to identify possible GW bursts: Exponential Gaussian 
Correlator (EGC), $\Omega$-pipeline ($\Omega$), and coherent WaveBurst (cWB). 
We analyze data from time intervals when at least two detectors were operating 
in coincidence.  Altogether, eight networks, or sets of detectors, operating 
during mutually exclusive time periods are analyzed by at least one algorithm.
Table~\ref{tab:data} shows the time available for analysis (``live time'')
for the different
network configurations after application of category 1 and 2 DQFs.
The actual times searched by each algorithm for each network
(``observation times'') reflect
details of the algorithms, such as the smallest analyzable data block,
as well as choices about which networks are most suitable for each algorithm.
The three- and two-detector
network configurations not shown in Table~\ref{tab:data} have negligible
live time and are not considered in this search.
\begin{table}[htbp]
\begin{tabular}{|c|c|c|c|c|} 
\hline
  network    &   live time   & cWB    & $\Omega$ & EGC  \\ 
\hline
 H1H2L1V1    &   68.9        & 68.2   & 68.7     & 66.6 \\
\hline
 H1H2L1      &   124.6       & 123.2  & 123.4    & 16.5 \\
 H1H2V1      &   15.8        & 15.7   & 15.1     & 15.3 \\ 
 H1L1V1      &   4.5         & 4.2    & -        & 4.4  \\
\hline
 H1H2        &   35.4        & 35.2   & 34.8     & -    \\
 H1L1        &   7.2         & 5.9   & -        & -    \\
 L1V1        &   6.4         &  -     & 6.3     & -    \\ 
 H2L1        &   3.8         & 3.5    & -        & -    \\
 \hline
\end{tabular}
\caption{\label{tab:data} Exclusive live time in days for each detector network
configuration after category 2 DQFs (second column) and the observation time
analyzed by each of the search algorithms (last three columns). 
The cWB algorithm did not process the L1V1 network because the 
coherent likelihood regulator used in this analysis was suboptimal
for two detectors with very different orientations.
Omega used a coherent combination of H1 and H2 as an effective detector
and thus analyzed networks either with both or with neither.
EGC analyzed only data with three or more interferometers during the
part of the run when Virgo was operational.
}
\end{table}

LIGO and GEO\,600 data are sampled at 16384~Hz, yielding a maximum
bandwidth of 8192~Hz, while Virgo data are sampled at 20000~Hz.
Because of the large calibration uncertainties at high frequency, 
only data below 6000~Hz are used in the search. Also, because of 
high seismic noise, the frequency band below 50~Hz is excluded from 
the analysis.  Furthermore, the EGC search was limited to the
300--5000~Hz band over which Virgo's sensitivity was comparable to
LIGO's.
 In Section~\ref{sec:systematics} we describe the influence of 
the calibration uncertainties on the results of the search.

\subsection{Vetoes}

After gravitational-wave candidate events are identified by the search
algorithms, they are subject to additional ``veto'' conditions to exclude 
events occurring within certain time intervals. These vetoes are based on 
statistical correlations between transients in the GW channel (data stream) 
and the environmental and interferometric auxiliary channels. 

We define vetoes for \sfyt/VSR1 following the 
approach used for \sfyo~\cite{S5y1Burst}.  
More details are given in Appendix~\ref{sec:vetoes}.

\subsection{Background Estimation and Tuning}

To estimate the significance of candidate GW events, and to optimize 
event selection cuts, we need to measure the distribution of events 
due to background noise.  With a multi-detector analysis one can create a 
sample of background noise events and study its statistical properties.
These samples are created by time-shifting data of one or more detectors 
with respect to the others by ``un-physical'' time delays ({\it i.e.\/} much 
larger than the maximum time-of-flight of a GW signal between the detectors).
Shifts are typically in the range from $\sim$1 s to a few minutes. 
Any triggers that are coincident in the time-shifted data cannot be due to a 
true gravitational-wave signal; these coincidences therefore sample the noise 
background.
Background estimation is done separately for each algorithm and network 
combination, using hundreds to thousands of shifts.
To take into account possible correlated noise transients in the H1 and H2 
detectors, which share a common environment and vacuum system, no time-shifts 
are introduced between these detectors for any network combination including 
another detector.

The shifted and unshifted data are analyzed identically.
A portion of the background events are used together with 
simulations (see below) to tune the search thresholds and 
selection cuts; the remainder is used to estimate the 
significance of any candidate events in the unshifted data 
after the final application of the selection thresholds.
All tuning is done purely on the time shifted data and 
simulations prior to examining the unshifted data-set.
This ``blind'' tuning avoid any biases in our candidate selection. 
The final event thresholds are determined by optimizing
the detection efficiency of the algorithms at a fixed false alarm rate.

\subsection{Hardware and software injections}

At pseudo-random times during the run, 
simulated burst signals were injected (added) into the interferometers
by sending pre-calculated waveforms to the mirror position control
system. These ``hardware injections'' provided an
end-to-end verification of the detector 
instrumentation, the data acquisition system and the data analysis software.
The injection times were clearly marked in the
data with a DQF.
Most of hardware injections were incoherent, {\it i.e.}, performed into a single detector
with no coincident injection into the other detectors. Some injections
were performed coherently by taking into account a simulated source location in the sky
and the angle-dependent sensitivity of the detectors to the two wave
polarization states.

In addition to the flagged injections, a ``blind injection challenge''
was undertaken in which a small number (possibly zero) of coherent 
hardware injections were performed {\em without} being marked by a DQF.  
Information about these blind injections (including whether the number 
was nonzero) was hidden from the data analysis teams during the search, 
and revealed only afterward.  This challenge was
intended to test our data analysis procedures and decision
processes for evaluating any candidate events that might be found by the search algorithms. 

To determine the sensitivity of our search to gravitational waves, 
and to guide the tuning of selection cuts, we repeatedly re-analyze the 
data with simulated signals injected in software.  
The same injections are analyzed by all three analysis pipelines.
See Section~\ref{sec:simulations} for more details.

\section{Search algorithms}
\label{sec:pipeline}

Anticipated sources of gravitational wave bursts are usually not understood
well enough to generate waveforms accurate and precise enough for matched
filtering of generic signals.
While some sources of GW bursts are being modeled with increasing success, 
the results tend to be highly dependent on physical parameters which 
may span a large parameter space. 
Indeed, some burst signals, such the white-noise burst from turbulent 
convection in a core-collapse supernova, are stochastic in nature and 
so are inherently not templatable. 
Therefore usually more robust
excess-power algorithms~\cite{anderson01,chatterjiThesis,finn04,waveburst04} 
are employed in burst searches.  By measuring power in the 
data as a function of time and frequency,
one may identify regions where the power is not consistent
with the anticipated fluctuations of detector noise. 
To distinguish environmental and instrumental transients from true 
GW signals, a multi-detector analysis approach is normally used, 
in which the event must be seen in more than one detector to be 
considered a candidate GW.

The simplest multi-detector analysis strategy is to require that the 
events identified in the individual detectors are coincident in time. 
The time coincidence window which should be chosen to take into account 
the possible time delays of a GW signal arriving at different sites, 
calibration and algorithmic timing biases, and possible signal model 
dependencies. Time coincidence can be augmented by requiring also 
an overlap in frequency.  One such time-frequency coincidence method 
used in this search is the EGC algorithm~\cite{beauville07} (see also 
Appendix~\ref{sec:EGC}).  It estimates the signal-to-noise ratio (SNR) 
$\rho_k$ in each detector $k$ and uses the combined SNR 
$\rho_{\text{comb}}=\sqrt{\sum_k{\rho^2_k}}$ to rank candidate events.  

A modification of the time-frequency coincidence approach is used in 
the $\Omega$ search algorithm~\cite{multiresolution} (also see 
Appendix~\ref{sec:omega}). In $\Omega$, the identification of the H1H2 
network events is improved by coherently combining the H1 and H2 data 
to form a single pseudo-detector data stream H$_+$.  This algorithm 
takes an advantage of the fact that the co-located and co-aligned H1 
and H2 detectors have identical responses to a GW signal.  The 
performance of the $\Omega$ algorithm is further enhanced by requiring 
that no significant power is left in the H1$-$H2 null stream, H$_{-}$, 
where GW signals cancel.  This veto condition helps to reduce 
the false alarm rate due to random coincidences of noise transients, 
which typically leave significant power in the null stream.  Network 
events identified by $\Omega$ are characterized by the strength 
$Z=\rho^2/2$ of the individual detector events, and by the correlated
H1H2 energy $\Zcor$.

A different network analysis approach is used in the cWB search 
algorithm~\cite{klimenko08} (see also~\cite{S5y1Burst} and 
Appendix~\ref{sec:cWB}).
The cWB algorithm performs a least-squares fit of a common GW signal to the data 
from the different detectors using the constrained likelihood 
method~\cite{klimenko05}.  The results of the fit are estimates of 
the $h_+$ and $h_{\times}$ waveforms, the most probable source location in 
the sky, and various likelihood statistics used in the cWB selection cuts.
One of these is the maximum likelihood ratio $L_m$, which is an 
estimator of the total SNR  
detected in the network.  A part of the $L_m$ statistic depending on 
pairwise combinations of the detectors is used to construct the network 
correlated amplitude $\eta$, which measures the degree of correlation 
between the detectors. 
Random coincidences of noise transients typically give low values of $\eta$, 
making this statistic useful for background rejection.
The contribution of each detector to the total SNR is weighted depending on the variance of the
noise and angular sensitivity of the detectors. The algorithm automatically marginalizes a detector
with either elevated noise or unfavorable antenna patterns, so that
it does not limit the sensitivity 
of the network. 

\section{Simulated signals and efficiencies}\label{sec:simulations}

\begin{figure}
\begin{center}
\mbox{\includegraphics*[width=0.5\textwidth]{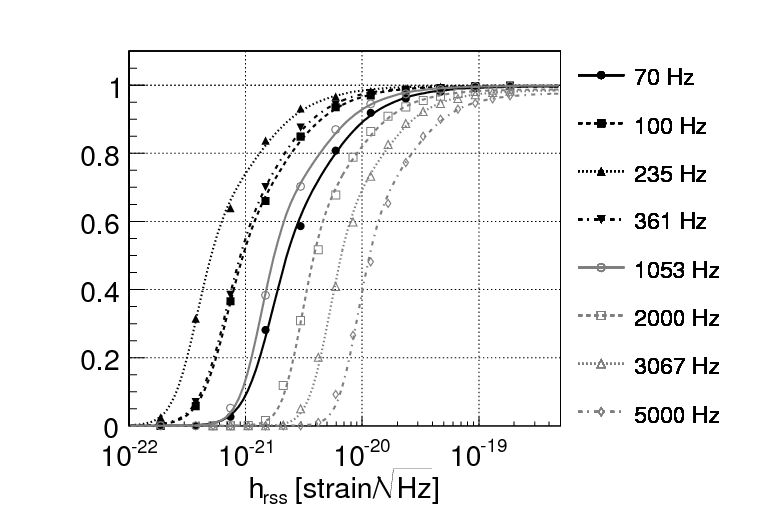}}
\mbox{\includegraphics*[width=0.5\textwidth]{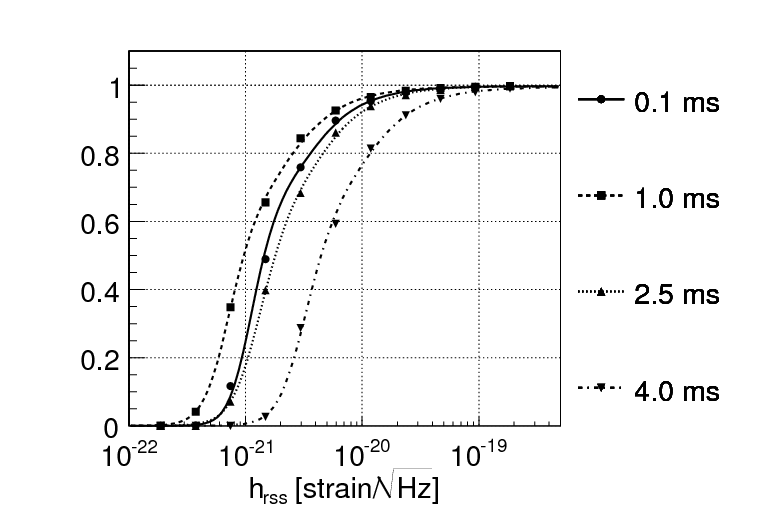}}
\mbox{\includegraphics*[width=0.5\textwidth]{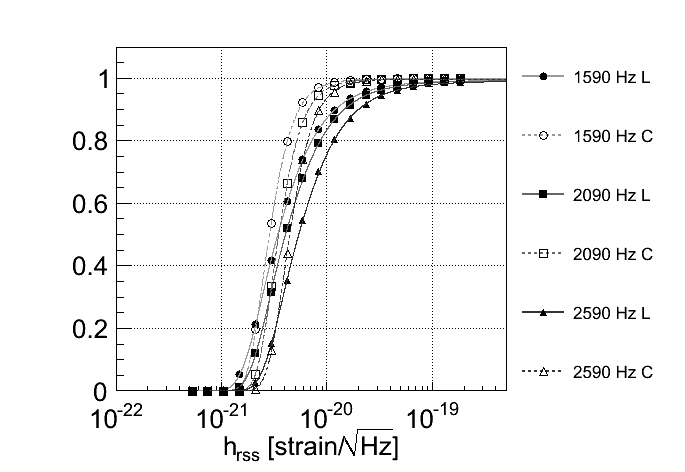}}
\caption{Efficiency for selected waveforms as a function of signal 
amplitude $h_{\text{rss}}$ for the logical OR of the H1H2L1V1, 
H1H2L1, and H1H2 networks.  Top: sine-Gaussians with $Q=9$ and 
central frequency spanning between $70$ and $5000$ Hz.  Middle: 
Gaussians with $\tau$ between $0.1$ and $4.0$ ms.  Bottom: linearly 
(L) and circularly (C) polarized ringdowns with $\tau=200$ ms and 
frequencies between $1590$ and $2590$ Hz. 
}
\label{fig:eff-curves}
\end{center}
\end{figure}

The detection efficiencies of the search algorithms depend on the
network configuration, the selection cuts used in the analysis, and
the GW morphologies which may span a wide range of signal durations,
frequencies and amplitudes.  To evaluate the sensitivity of the search
and verify that the search algorithms do not have a strong model
dependency, we use several sets of ad-hoc waveforms.  These include
\begin{description}
\item{Sine-Gaussian} waveforms:
\begin{eqnarray} 
h_+(t) & = & h_0 \sin(2\pi f_0 t)\exp[-(2\pi f_0 t)^2/2Q^2], \\
h_\times(t) & = & 0 \, .
\end{eqnarray}
We use a discrete set of central frequencies $f_0$ from 70 Hz to 6000 Hz 
and quality factors $Q$ of 3, 9, and 100; 
see Table~\ref{table:SGQ9} and Fig.~\ref{fig:eff-curves} (top). 
The amplitude factor $h_0$ is varied to simulate GWs with different
strain amplitudes.
For definition of the polarizations, see Eq.~(\ref{hdet}) and text below it.
\item{Gaussian} waveforms:
\begin{eqnarray} 
h_+(t) & = & h_0 \exp(-t^2/\tau^2 ), \\
h_\times(t) & = & 0 \, ,
\end{eqnarray}
where the duration parameter $\tau$ is chosen to be one of 
(0.1, 1.0, 2.5, 4.0) ms; see
Fig.~\ref{fig:eff-curves} (middle).  
\item{Harmonic ringdown} signals:
\begin{eqnarray} 
h_+(t) & = & h_{0,+} \cos(2\pi f_0 t)\exp[-t/{\tau}], \quad t>1/(4f_0), \\
h_\times(t) & = & h_{0,\times} \sin(2\pi f_0 t)\exp[-t/{\tau}], \quad t>0 \, .
\end{eqnarray}
We use several central frequencies $f_0$ from 1590 Hz to 3067 Hz, one 
long decay time, $\tau=200$~ms, and two short decay times, 1 ms and 0.65 ms;
see Table~\ref{table:RD} and Fig.~\ref{fig:eff-curves} (bottom). 
Two polarization states are used: circular ($h_{0,+} = h_{0,\times}$), 
and linear ($h_{0,+}=0$).
The quarter-cycle delay in $h_+$ is to avoid starting the waveform with a
large jump.
\item{Band-limited white noise} signals: 

These are bursts of Gaussian noise which are white over 
a frequency band $[f_{\text{low}}, f_{\text{low}}+\Delta{f}]$ and 
which have a Gaussian time profile with standard deviation decay time 
$\tau$; see Table~\ref{table:WNB}.  These signals are unpolarized in 
the sense that the two polarizations $h_+$ and $h_\times$ have equal 
RMS amplitudes and are uncorrelated with each other.
\end{description} 
The strengths of the ad hoc waveform injections are characterized by 
the root-square-sum 
amplitude $h_{\text{rss}}$,
\begin{equation}
h_{\text{rss}} = \sqrt{\int_{-\infty}^{+\infty} \!\!\! dt \, \left( |h_+(t)|^2 + |h_{\times}(t)|^2 \right)}.
\end{equation}

The parameters of these waveforms are selected to coarsely cover the
frequency range of the search from $\sim$50 Hz to $\sim$6 kHz, and duration of
signals up to a few hundreds of milliseconds.  The Gaussian,
sine-Gaussian and ringdown waveforms explore the space of GW signals
with small time-frequency volume, while the white noise bursts explore
the space of GW signals with relatively large time-frequency volume.
Although the simulated waveforms are not physical, they may be similar
to some waveforms produced by astrophysical sources.  For example, the
sine-Gaussian waveforms with few cycles are qualitatively similar to
signals produced by the mergers of two black holes~\cite{bbh}.  The
long-timescale ringdowns are similar to signals predicted for
excitation of neutron-star fundamental modes~\cite{benhar04}.  Some
stellar collapse and core-collapse supernova models predict signals
that resemble short ringdown waveforms (in the case of a
rapidly rotating progenitor star) or band-limited white-noise
waveforms with random polarizations. In the context of the recently
proposed acoustic mechanism for core-collapse supernova explosions,
quasi-periodic signals of $\gtrsim$500 ms duration have been proposed
\cite{ott2008}.

To test the range for detection of gravitational waves from neutron
star collapse, two waveforms were taken from simulations by Baiotti
{\it et al.}~\cite{SNwave}, who modeled neutron star gravitational
collapse to a black hole and the subsequent ringdown of the black hole
using collapsing polytropes deformed by rotation.  The models whose
waveform we chose were D1, a nearly spherical 1.67 $M_\odot$ neutron
star, and D4, a 1.86 $M_\odot$ neutron star that is maximally deformed
at the time of its collapse into a black hole.  These two specific
waveforms represent the extremes of the parameter space in mass and
spin considered in ~\cite{SNwave}. They are linearly polarized
($h_\times=0$), with the waveform amplitude varying with the
inclination angle $\iota$ (between the wave propagation vector and symmetry
axis of the source) as $\sin^2\iota$.

The simulated detector responses $h_{\text{det}}$ are constructed as 
\begin{equation}
\label{hdet}
h_{\text{det}} = F_+(\theta, \phi, \psi) h_+ + F_{\times}(\theta, \phi, \psi) h_{\times} \, .
\end{equation}
Here $F_+$ and $F_{\times}$ are the detector antenna patterns, which 
depend on the direction to the source ($\theta,\phi$) and the polarization 
angle $\psi$.
(The latter is defined as in Appendix~B of~\cite{anderson01}.)
These parameters are chosen randomly for each injection. 
The sky direction is isotropically distributed, and the random polarization 
angle is uniformly distributed on $[0,\pi)$.  The injections are distributed 
uniformly in time across the \sfyt/VSR1 run, with an average separation of 100 s.
Note that for the ad-hoc waveforms no $\iota$ is used.

\begin{figure}
\begin{center}
\mbox{\includegraphics*[width=0.5\textwidth]{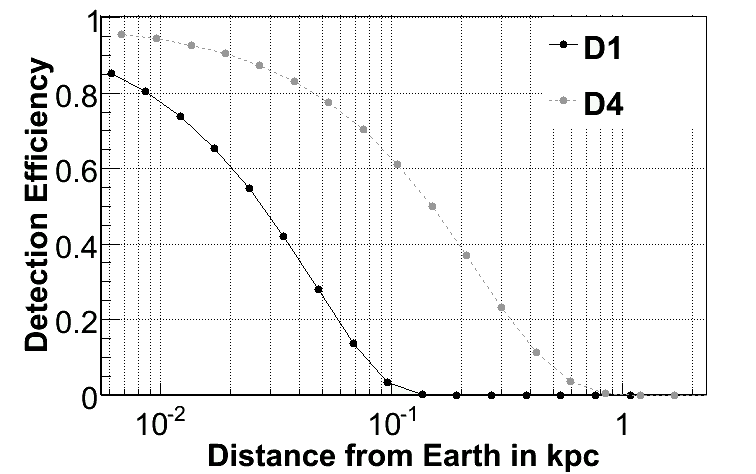}}
\caption{Efficiency of the H1H2L1V1 network as a function of distance 
for the D1 and D4 waveforms of Baiotti {\it et al.}~\cite{SNwave} 
predicted by polytropic general-relativistic 
models of neutron star collapse.  These efficiencies assume random sky 
location, polarization and inclination angle.}
\label{fig:D1D4range}
\end{center}
\end{figure}

The detection efficiency after application of all selection cuts was
determined for each waveform type.  All waveforms were evaluated using
cWB, while subsets were evaluated using $\Omega$ and EGC, due mainly
to the limited frequency bands covered by those algorithms as they
were used in this search (48--2048~Hz and 300--5000~Hz, respectively).
Figure~\ref{fig:eff-curves} shows the combined efficiency curves
for selected sine-Gaussian, Gaussian and ringdown simulated signals 
as a function of the $h_{\text{rss}}$ amplitude.
Figure~\ref{fig:D1D4range} shows the detection efficiency for the 
astrophysical signals D1 and D4 as a function of the distance to the source.

Each efficiency curve is fitted with an empirical function and 
the injection amplitude for which that function equals 50\% is determined. 
This quantity, $h_{\text{rss}}^{50\%}$, is a convenient characterization of 
the sensitivity of the search to that waveform morphology.  
Tables~\ref{table:SGQ9}, \ref{table:RD}, and \ref{table:WNB} 
summarize the sensitivity of the search to the sine-Gaussian, ringdown, and 
band-limited white noise burst signals.
Where possible, we also calculate the sensitivity of the logical OR  
of the cWB and $\Omega$ algorithms (since those two are used for the
upper limit calculation as described in Sec.~\ref{sec:results}), and
for the appropriately weighted combination of all networks (some of
which are less sensitive) contributing to the total observation time.
In general, the efficiency of the combination of the search algorithms 
is slightly more sensitive than the individual algorithms. 

\begin{table}
  \centering
  \begin{tabular}{cc|ccc|c|cc}\hline\hline
$f_0$ & $Q$  & \multicolumn{4}{c|}{H1H2L1V1, $h_{\text{rss}}^{50\%}$ }  & \multicolumn{2}{c}{all networks }\\ 
 $[Hz]$ &   &  cWB   &  $\Omega$    & EGC & cWB or $\Omega$ & $h_{\text{rss}}^{50\%}$ & $h_{\text{rss}}^{90\%}$\\ \hline
  70    &  3       &  17.9  &  26.7  & -    &  17.6 &  20.4 & 96.6\\
  70    &  9       &  20.6  &  34.4  & -    &  20.6 &  25.0 & 120\\
  70    &  100     &  20.5  &  35.0  & -    &  20.0 &  25.1 & 121\\
  100   &  9       &   9.2  &  14.1  & -    &   9.1 &  10.6 & 49.7\\
  153   &  9       &   6.0  &   9.1  & -    &   6.0 &   6.5 & 29.3\\
  235   &  3       &   6.5  &   6.6  & -    &   5.9 &   6.1 & 28.8\\
  235   &  9       &   6.4  &   5.8  & -    &   5.6 &   5.6 & 26.8\\
  235   &  100     &   6.5  &   6.7  & -    &   6.2 &   6.0 & 26.1\\
  361   &  9       &  10.5  &  10.2  & 60.1 &   9.5 &  10.0 & 42.0\\
  554   &  9       &  11.1  &  10.5  & 18.8 &   9.9 &  10.9 & 47.1\\
  849   &  3       &  19.2  &  15.8  & 30.0 &  15.3 &  15.8 & 73.8\\
  849   &  9       &  17.7  &  15.3  & 28.5 &  14.6 &  15.8 & 71.5\\
  849   &  100     &  16.0  &  16.2  & 31.3 &  14.5 &  15.3 & 66.7\\
  1053  &  9       &  22.4  &  19.0  & 33.8 &  18.3 &  19.4 & 86.9\\
  1304  &  9       &  28.1  &  23.6  & 41.0 &  22.6 &  24.7 & 115\\
  1451  &  9       &  28.6  &   -    & 43.3 &  28.6 &  30.2 & 119\\   
  1615  &  3       &  39.6  &  32.1  & 48.4 &  31.7 &  33.8 & 146\\
  1615  &  9       &  33.7  &  28.1  & 51.1 &  27.3 &  29.5 & 138\\
  1615  &  100     &  29.6  &  30.6  & 53.8 &  27.6 &  28.6 & 126\\
  1797  &  9       &  36.5  &   -    & 57.8 &  36.5 &  38.3 &  146\\   
  2000  &  3       &  42.6  &   -    &  -   &  42.6 & 47.1  & 191\\
  2000  &  9       &  40.6  &   -    & 58.7 &  40.6 & 44.0  & 177\\
  2000  &  100     &  34.9  &   -    &  -   &  34.9 & 38.4  & 153\\   
  2226  &  9       &  46.0  &   -    & 68.6 &  46.0 & 51.1  & 187\\
  2477  &  3       &  61.9  &   -    & -    &       61.9  &       65.6   &       262\\
  2477  &  9       &  53.5  &   -    & 76.7 &       53.5  &       56.1  &       206\\
  2477  &  100     &  44.5  &   -    & -    &       44.5  &       48.9  &       201\\   
  2756  &  9       &  60.2  &   -    & 82.2 &       60.2  &       64.4  &       248\\
  3067  &  3       &  86.9  &   -    & -    &       86.9  &       87.0  &       343 \\ 
  3067  &  9       &  69.0  &   -    & 96.6 &       69.0  &       75.0  &       286\\
  3067  &  100     &  55.4  &   -    & -    &       55.4   &       61.1  &       273 \\  
  3413  &  9       &  75.9  &   -    & 108 &       75.9  &       82.9  &       323\\
  3799  &  9       &  89.0  &   -    & 116  &       89.0  &       97.7  &       386\\
  4225  &  9       &  109   &   -    & 138  &       109  &       115  &       575\\
  5000  &  3       &  207   &   -    & -    &       207  &       187  &       1160 \\   
  5000  &  9       &  126   &   -    & 155  &       126  &       130  &       612\\
  5000  &  100     &  84.7  &   -    & -    &       84.7  &       100  &       480 \\   
  6000  &  9       &  182   &   -    & -    &       182  &       196  &       893\\

\hline\hline
\end{tabular}
\caption{Values of $h_{\text{rss}}^{50\%}$ and $h_{\text{rss}}^{90\%}$ (for 
50\% and 90\% detection efficiency),
in units of $10^{-22} \, {\rm Hz}^{-1/2}$,
for sine-Gaussian waveforms with the central frequency $f_0$ and quality factor $Q$. 
Three columns in the middle are the $h_{\text{rss}}^{50\%}$ measured with the individual 
search algorithms for the H1H2L1V1 network. The next column is the $h_{\text{rss}}^{50\%}$ 
of the logical OR of the cWB and $\Omega$ algorithms for the H1H2L1V1 network. 
The last two columns  are the $h_{\text{rss}}^{50\%}$ and the $h_{\text{rss}}^{90\%}$ of the logical OR of the algorithms
and networks (H1H2L1V1 or H1H2L1 or H1H2).
All $h_{\text{rss}}$  values take into account statistical and systematic
uncertainties as explained in Sec.~\ref{sec:systematics}.
} 
\label{table:SGQ9}
\end{table}

\begin{table}
\centering
\begin{tabular}{cc|cc|cc}\hline\hline
$f$&  $\tau$  &\multicolumn{2}{c|}{all networks, $h_{\text{rss}}^{50\%}$ }  & \multicolumn{2}{c}{all networks, $h_{\text{rss}}^{90\%}$} \\ 
$[Hz]$ & $[ms]$  &   Lin. &   Circ. & Lin. &  Circ. \\ \hline
1590    &       200   &       34.7 &       30.0 &       131 &       60.0\\
2000    &       1.0      &       49.5 &       43.8 &       155 &       81.1\\
2090    &       200   &       43.3 &       36.5 &       155 &       72.9\\
2590    &       200   &       58.6 &       46.0 &       229 &       88.8\\
3067    &       0.65      &       88.2 &       73.3 &       369 &       142\\
\hline\hline
\end{tabular}
\caption{Values of $h_{\text{rss}}^{50\%}$ and $h_{\text{rss}}^{90\%}$ 
(for 50\% and 90\% detection efficiency using cWB),
in units of $10^{-22} \, {\rm Hz}^{-1/2}$,
for linearly and circularly polarized ringdowns characterized by parameters $f$ and $\tau$.
All $h_{\text{rss}}$ values 
take into account statistical and systematic
uncertainties as explained in Sec.~\ref{sec:systematics}.
}
\label{table:RD}
\end{table}

\begin{table}
\centering
\begin{tabular}{ccc|cc|c|cc}\hline\hline
$f_{\text{low}}$& $\Delta{f}$ & $\tau$  &\multicolumn{3}{c|}{H1H2L1V1, $h_{\text{rss}}^{50\%}$ }  & \multicolumn{2}{c}{all networks } \\ 
$[Hz]$ & $[Hz]$ & $[ms]$  &   cWB &   $\Omega$ & cWB or $\Omega$ &  $h_{\text{rss}}^{50\%}$ & $h_{\text{rss}}^{90\%}$ \\ \hline
100    & 100    & 0.1     &  7.6   &  13.6  &  7.6   &  8.4   &  19.6 \\
250    & 100    & 0.1     &  9.1   &  10.2  &  8.8   &  8.6   &   18.7\\
1000   & 10     & 0.1     & 20.9   &  28.6  & 21.0   & 21.8   &  52.6 \\
1000   & 1000   & 0.01    & 36.8   &  38.2  & 35.0   & 36.3   &  74.7 \\
1000   & 1000   & 0.1     & 60.3   &  81.7  & 60.7   & 63.5   &  140 \\
2000   & 100   & 0.1    & 40.4   &  -     & 40.4   & 44.1    &       94.4  \\
2000   & 1000   & 0.01    & 60.7   &  -     & 60.7   & 62.4     &       128  \\
3500   & 100   & 0.1    & 74.3   &  -     & 74.3   & 84.8  &       182    \\
3500   & 1000   & 0.01    & 103   &  -     & 103   & 109   &       224   \\
5000   & 100   & 0.1    & 101  &  -     & 101  & 115  &       255   \\
5000   & 1000   & 0.01    & 152  &  -     & 152  & 144  &       342   \\

\hline\hline
\end{tabular}
\caption{Values of $h_{\text{rss}}^{50\%}$ and $h_{\text{rss}}^{90\%}$ 
(for 50\% and 90\% detection efficiency), in units of 
$10^{-22} \, {\rm Hz}^{-1/2}$, for band-limited noise waveforms 
characterized by parameters $f_{\text{low}}$, $\Delta{f}$, and $\tau$.
Two columns in the middle are the $h_{\text{rss}}^{50\%}$ for the 
individual search algorithms for the H1H2L1V1 network. The next column 
is the $h_{\text{rss}}^{50\%}$ of the logical OR of the cWB and $\Omega$ 
algorithms for the H1H2L1V1 network. The last two columns  are the 
$h_{\text{rss}}^{50\%}$ and the $h_{\text{rss}}^{90\%}$ of the logical 
OR of the algorithms and networks (H1H2L1V1 or H1H2L1 or H1H2).
All $h_{\text{rss}}$  values take into account statistical and systematic
uncertainties as explained in Sec.~\ref{sec:systematics}.
}
\label{table:WNB}
\end{table}

\section{Uncertainties}\label{sec:systematics}

The amplitude sensitivities presented in this paper, {\it i.e.}\ the
$h_{\text{rss}}$ values at 50\% and 90\% efficiency, have been
adjusted upward to conservatively reflect statistical and systematic
uncertainties.
The statistical uncertainty arises from the limited number of
simulated signals used in the efficiency curve fit, and is typically a
few percent.
The dominant source of systematic uncertainty
comes from the amplitude calibration:
the single detector amplitude calibration uncertainties is typically of order 10\%.   
Neglegible effects are due to phase and timing uncertainties.  

The amplitude calibration of the interferometers is less accurate at high 
frequencies than at low frequencies, and
therefore two different approaches to handling calibration uncertainties are used in 
the \sfyt/VSR1 search. In the frequency band below 2 kHz, we use  
the procedure established for \sfyo~\citep{S5y1BurstHF}.  We combine the amplitude 
uncertainties from each interferometer into a single uncertainty by calculating 
a combined root-sum-square
amplitude SNR and propagating the individual uncertainties
assuming each error is independent: as a conservative result, the detection efficiencies are 
rigidly shifted towards higher $h_{\text{rss}}$ by 11.1\%.
In the frequency band above 2 kHz, a new methodology, based on MonteCarlo simulations 
 has been adopted to marginalize over calibration uncertainties: basically, we inject signals whose amplitude has been jittered according to 
the calibration uncertainties.
The effect of miscalibration resulted in the increase of the combined $h_{\text{rss}}^{50\%}$ 
by 3 \% to 14\%, depending mainly on the central frequency of the injected signals.
\section{Search Results}\label{sec:results}

In Section~\ref{sec:overview} we described the main steps in our 
search for gravitational-wave bursts.  In the search all 
analysis cuts and thresholds are set in a blind way, using 
time-shifted (background) and simulation data.  The blind cuts are 
set to yield a false-alarm rate of approximately 0.05 events or less
over the observation time of each search algorithm, 
network configuration and target frequency band.  
Here we describe the results.

\subsection{Candidate events}
\label{ss:candidates}

After these cuts are fixed, the unshifted events are examined and the 
various analysis cuts, DQFs, and vetoes are applied.  Any surviving 
events are considered as candidate gravitational-wave events and 
subject to further examination.
The purpose of this additional step is to go beyond the binary
decision of the initial cuts and evaluate additional information 
about the events which may reveal their origin. This ranges from 
``sanity checks'' to deeper investigations on the background of 
the observatory, detector performances, environmental disturbances 
and candidate signal characteristics.

\begin{figure}
\begin{center}
\includegraphics*[width=0.5\textwidth]{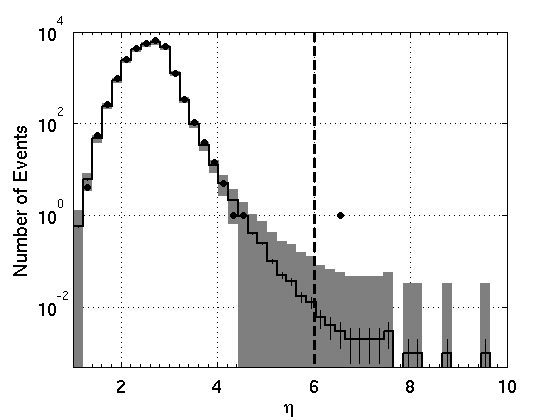} 
\includegraphics*[width=0.5\textwidth]{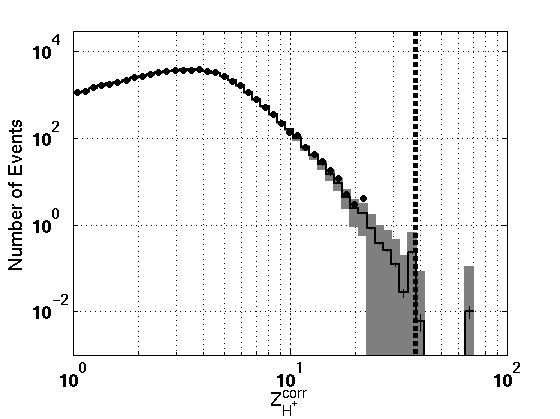}\\
\caption{Distribution of background (solid line) and foreground
(solid dots) events from the search below 200~Hz in the H1H2L1 
network, after application of category 2 data quality and vetoes: 
cWB (top), $\Omega$ (bottom).
The event-strength figures of merit on the horizontal axes are
defined in the appendices on the search algorithms.
The small error bars on the solid 
line are the 1 $\sigma$ statistical uncertainty on the estimated 
background, while the wider gray belt represents the expected 
root-mean-square statistical fluctuations on the number of background 
events in the foreground sample.  The loudest foreground event on the 
top plot is the only event that survived the blind detection cuts of 
this search, shown as vertical dashed lines.  This event was later 
revealed to have been a blind injection.
} 
\label{fig:cWBrateEQ}
\end{center}
\end{figure}

Examining the unshifted data, we found one foreground event among all 
the different search algorithms and detector combinations that survives 
the blind selection cuts. It was produced by cWB during a time when all 
five detectors were operating simultaneously. As the possible first
detection of a gravitational-wave signal, this event was examined in 
great detail according to our follow-up checklist.  We found no evident 
problem with the instruments or data, and no environmental or instrumental
disturbance detected by the auxiliary channels.  The event was detected at 
a frequency of 110~Hz, where all detectors are quite non-stationary,
and where both the GEO\,600 and Virgo detectors had poorer sensitivity 
(see Fig.~\ref{fig:Shh}).  Therefore, while the event was found in the 
H1H2L1V1 analysis, we also re-analyzed the data using cWB and the H1H2L1 network. 
Figure~\ref{fig:cWBrateEQ} (top) shows the event above the blind selection 
cuts and the comparison with the measured H1H2L1 background of cWB in the 
frequency band below 200~Hz. 

\begin{figure}
\begin{center}
\includegraphics*[width=0.5\textwidth]{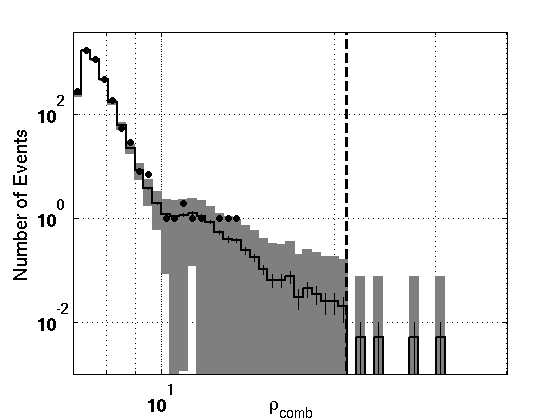}
\includegraphics*[width=0.5\textwidth]{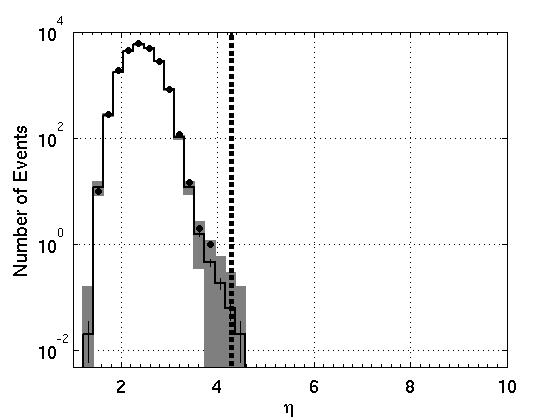}
\caption{Distribution of background (solid line) and foreground
(solid dots) H1H2L1V1 events after category 2 data quality and vetoes: 
EGC events in the frequency band 300--5000~Hz (top), 
cWB events in the frequency band 1200--6000~Hz (bottom). 
The event-strength figures of merit on the horizontal axes are
defined in the appendices on the search algorithms.
The small error bars on the solid line are the 1 $\sigma$ statistical 
uncertainty on the estimated background, while the wider gray belt 
represents the expected root-mean-square statistical fluctuations on 
the number of background events in the foreground sample.}
\label{fig:cWBrateHF}
\end{center}
\end{figure}

No foreground event passes the blind selection 
cuts in the $\Omega$ H1H2L1 analysis (see Figure~\ref{fig:cWBrateEQ} 
(bottom)); moreover, there is no visible excess of foreground events 
with respect to the expected background.  The cWB event is well 
within the tail of the $\Omega$ foreground and does not pass the 
final cut placed on correlated energy of the Hanford detectors.
Furthermore, the event is outside of the frequency band (300-5000~Hz) 
processed by the EGC algorithm.  Figure~\ref{fig:cWBrateHF} (top) shows 
the corresponding EGC foreground and background distributions for the 
H1H2L1V1 network.  
For comparison, Figure~\ref{fig:cWBrateHF} (bottom) shows similar 
distributions from cWB, with no indication of any excess of events 
in the frequency band 1200--6000~Hz.

To better estimate the significance of the surviving cWB event, we performed 
extensive background studies with cWB for the H1H2L1 network, accumulating a 
background sample with effective observation time of approximately 500 years.  
These studies indicate an expected false alarm rate for similar events 
of once per 43 years for the cWB algorithm and the H1H2L1 network. 
The statistical significance of the event must
take into account a ``trials factor'' arising from 
multiple analyses using different search algorithms, networks and 
frequency bands. Neglecting a small correlation among the backgrounds,  
this factor can be estimated by considering the total effective analyzed 
time of all the independent searches, which is 5.1 yr. The probability 
of observing one event at a background rate of once per 43 years or less 
in any of our searches is then on the order of $10\%$.  This probability 
was considered too high to exclude a possible accidental origin of this 
event, which was neither confirmed nor ruled out as a plausible GW signal.
This event was later revealed to be a hardware injection with $h_\mathrm{rss} =
1.0\times10^{-21}$~Hz$^{-1/2}$.
It was the only burst injection within the ``blind injection challenge.''
 Therefore it was removed from the analysis by the 
cleared injection data quality flag.  We can report that cWB recovered
the injection parameters and waveforms faithfully, and the exercise of treating
the event as a real GW candidate was a valuable learning experience.

Although no other outstanding foreground events were observed in the search, 
we have additionally examined events in the data set with relaxed selection cuts, 
namely, before applying category 3 DQFs and vetoes. In this set we find a total of 
three foreground events. One of these is produced by the EGC algorithm 
(0.16 expected from the background) and
the other two are from the $\Omega$-pipeline (1.4 expected).
While an exceptionally strong event in the enlarged
data set could, in principle, be judged to be a plausible GW signal,
none of these additional events is particularly compelling.
The EGC event occurred during a time of high seismic
noise and while the H2 interferometer was re-acquiring lock (and thus
could occasionally scatter light into the H1 detector), both of which
had been flagged as category 3 data quality conditions.
The $\Omega$-pipeline events fail the category 3 vetoes due to having corresponding 
glitches in H1 auxiliary channels. None of these three 
events passes the cWB selection cuts.
For these reasons, we do not consider any of them 
to be a plausible gravitational-wave candidate. Also, since these events
do not pass the predefined category 3 data quality and vetoes, they do not affect 
the calculation of the upper limits presented below.

\subsection{Upper limits}
\label{UL}

The \sfyt/VSR1 search includes the analysis of eight network 
configurations with three different algorithms.  We use the 
method presented in \cite{sutton2009} to combine the results 
of this search, together with the \sfyo~search \cite{S5y1Burst}, 
to set frequentist upper limits on the rate of burst events.   
Of the \sfyt~results, we include only the networks H1H2L1V1, 
H1H2L1 and H1H2, as the other networks have small observation times and their 
contribution to the upper limit would be marginal. Also, we decided 
{\it a priori} to use only the two algorithms which processed the
data from the full S5y2 run, namely cWB and $\Omega$.  (EGC only
analyzed data during the $\sim$5 months of the run when Virgo was
operational.)  We are left therefore with six analysis results
to combine with the \sfyo~results to produce a single upper limit on the rate 
of GW bursts for each of the signal morphologies tested.

As discussed in \cite{sutton2009}, the upper limit procedure 
combines the sets of surviving triggers according to which algorithm(s) 
and/or network detected any given trigger, and weights each trigger 
according to the detection efficiency of that algorithm and network combination.  
For the special case of no surviving events, the 90\% confidence
upper limit on the total event rate (assuming a Poisson distribution of
astrophysical events) reduces to 
\begin{equation}
R_{90\%}=\frac{2.3}{\epsilon_{tot}T} \, ,
\end{equation}
where $2.3 = -\log(1-0.9)$,
$\epsilon_{tot}$ is the detection efficiency of the union of all 
search algorithms and networks, and $T$ is the total observation time of the 
analyzed data sets.

\begin{figure}
\begin{center}
\mbox{\includegraphics*[width=0.55\textwidth]{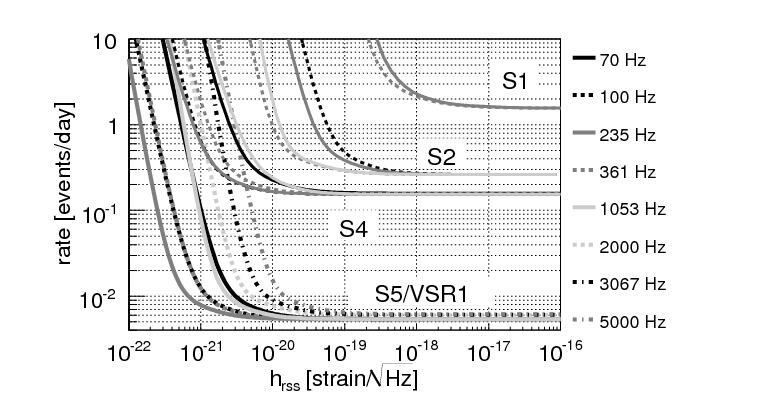}}
\mbox{\includegraphics*[width=0.55\textwidth]{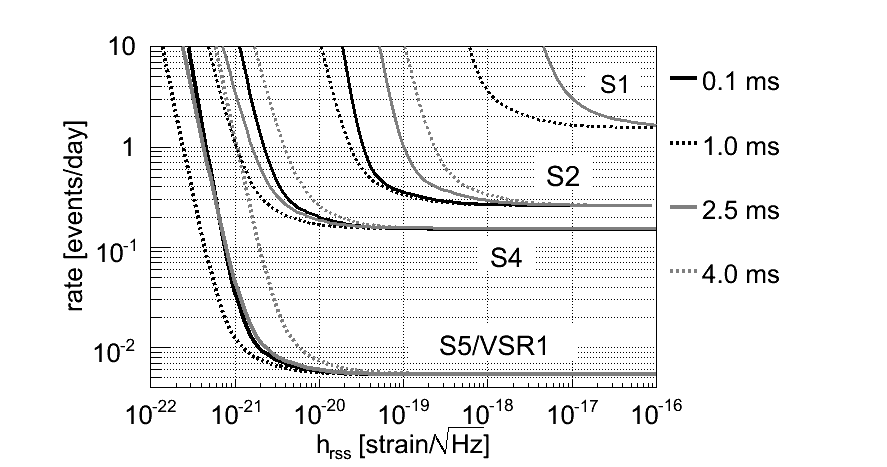}}
\caption{Selected exclusion diagrams showing the 90\% confidence rate limit as a function
of signal amplitude for Q=9 sine-Gaussian (top) and Gaussian (bottom) waveforms
for the results of the entire S5 and VSR1 runs (S5/VSR1) compared to the results 
reported previously (S1, S2, and S4).}
\label{fig:S5upperlimits}
\end{center}
\end{figure}

In the limit of strong signals in the frequency band below 2 kHz, 
the product $\epsilon_{tot}T$ is 224.0 days for \sfyo~and 205.3 days 
for \sfyt/VSR1. The combined rate limit for strong GW signals 
is thus $2.0\,\text{yr}^{-1}$.
For the search above 2 kHz, the rate limit for strong GW signals 
is $2.2\,\text{yr}^{-1}$.  This slightly weaker limit is due to the
fact that less data was analyzed in the \sfyo~high-frequency search 
than in the \sfyo low-frequency search (only 161.3 days of H1H2L1 data~\cite{S5y1BurstHF}).
Figure~\ref{fig:S5upperlimits} shows the combined rate limit as a function of amplitude 
for selected Gaussian and sine-Gaussian waveforms.

\begin{figure}
\begin{center}
\mbox{\includegraphics*[width=0.5\textwidth]{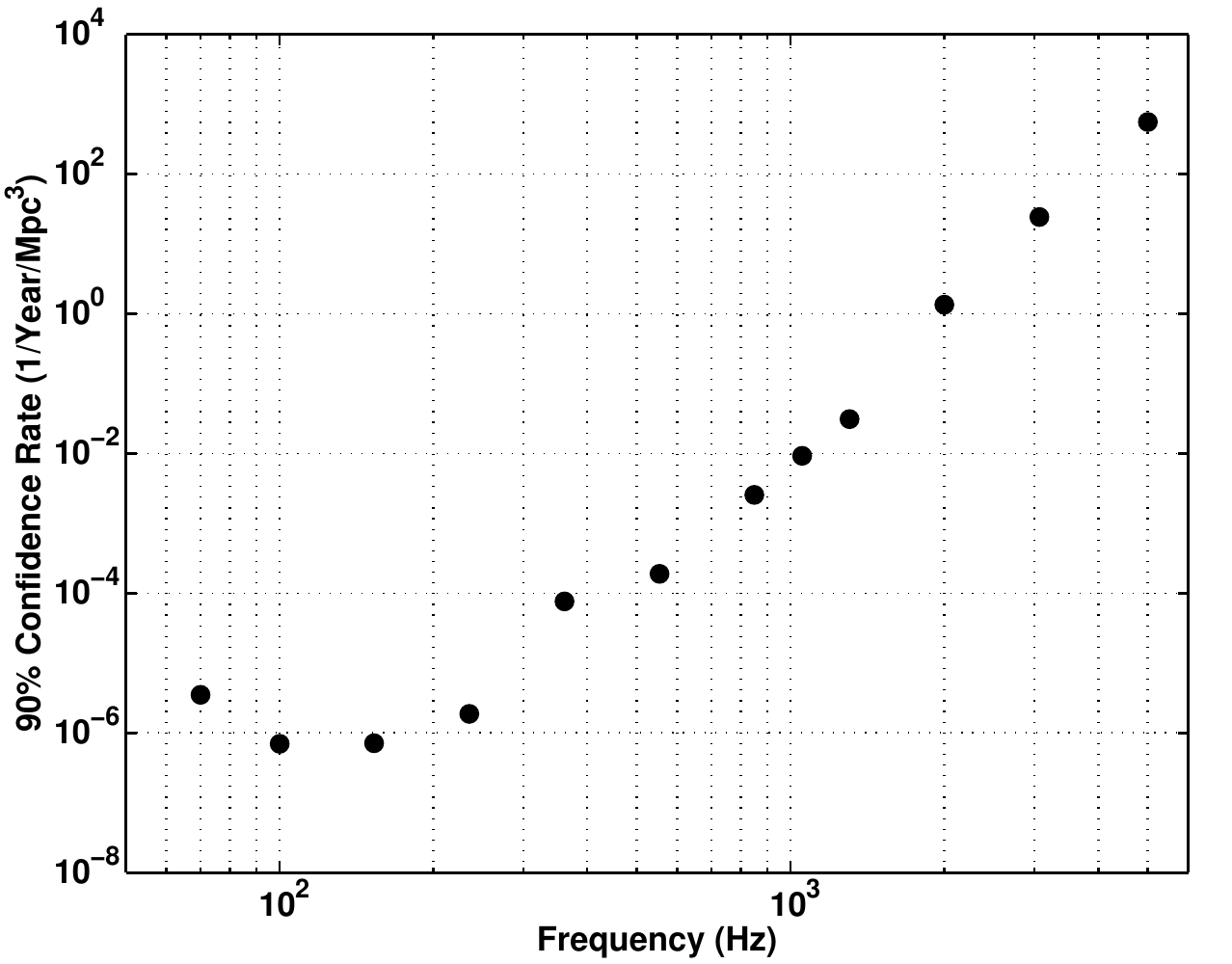}}
\caption{Rate limit per unit volume at the 90\% confidence level for a
linearly polarized
sine-Gaussian standard-candle with $E_{\text{GW}}=M_\odot{c^2}$.
}
\label{fig:isotropicUL}
\end{center}
\end{figure}

The results can also be interpreted as limits on the rate
density (number per time per volume) of GWBs assuming a standard-candle
source.  For example, given an isotropic distribution of sources with
amplitude $h_\mathrm{rss}$ at a fiducial distance $r_0$, and with rate density
${\cal R}$, the rate of GWBs at the Earth with amplitudes in 
the interval $[h,h+dh]$ is
\begin{equation}
dN = \frac{4\pi {\cal R} (h_\mathrm{rss} r_0)^3}{h^4} \, dh  \, .
\end{equation}
(Here we have neglected the inclination angle $\iota$; equivalently we can
take $h^2$ to be averaged over $\cos\iota$.)
The expected number of detections given the network efficiency $\epsilon(h)$
(for injections without any $\iota$ dependence) and the observation time $T$ is
\begin{eqnarray}
N_\mathrm{det} & = & T \int_0^\infty \!\! dh \, \left(\frac{dN}{dh}\right) \, \epsilon(h)
\nonumber \\
 & = &  4\pi {\cal R} T (h_\mathrm{rss} r_0)^3 \int_0^\infty  \!\! dh \, h^{-4} \epsilon(h) \, .
\end{eqnarray}
For linearly polarized signals distributed uniformly in $\cos\iota$, the
efficiency is the same with $h$ rescaled by a factor $\sin^2\iota$ divided by
that factor's appropriately averaged value $\sqrt{8/15}$.
Thus the above expression is multiplied by $\int_0^1 d\cos\iota (15/8)^{3/2}
\sin^6\iota \approx 1.17$.
The lack of detection candidates in the S5/VSR1 data set implies a 90\%
confidence upper limit on rate density ${\cal R}$ of
\begin{eqnarray}
{\cal R}_\mathrm{90\%} = \frac{2.0}{4\pi T (h_\mathrm{rss} r_0)^3 \int_0^\infty  \!\! dh \, h^{-4}
\epsilon(h)} \, .
\end{eqnarray}
Assuming that a
standard-candle source emits waves with energy $E_{\text{GW}}=M_\odot{c^2}$,
where $M_\odot$ is the solar mass, the product $h_\mathrm{rss} r_0$ is
\begin{eqnarray} \label{eq:hrVmass}
h_\mathrm{rss} r_0 = \sqrt{\frac{G M_\odot}{c}} (\pi f_0)^{-1}.
\end{eqnarray}
Figure~\ref{fig:isotropicUL} shows the rate density upper limits as a function of
frequency. This result can be interpreted in the following way: given a source
with a characteristic frequency $f$ and energy $E_{\text{GW}}=M{c^2}$,
the corresponding rate limit is 
${\cal R}_\mathrm{90\%}(f) (M_\odot / M)^{3/2}~\mathrm{yr}^{-1}\mathrm{Mpc}^{-3}$.
For example, for sources emitting at 150~Hz with $E_{\text{GW}}=0.01 M_\odot {c^2}$, 
the rate limit is approximately
$6\times 10^{-4}\mathrm{yr}^{-1}\mathrm{Mpc}^{-3}$.
The bump at $361$ Hz reflects the effect of the ``violin modes'' (resonant frequencies of the wires suspending the mirrors) on the sensitivity of the
detector.

\section{Summary and Discussion}\label{sec:summaries}

In this paper we present results of new all-sky untriggered searches for
gravitational wave bursts in data from the first Virgo science run (VSR1 in
2007) and the second year of the fifth LIGO science run (S5y2 in 2006--2007).
This data set represented the first long-term operation of a worldwide network
of interferometers of similar performance at three different sites.
Data quality and analysis algorithms have improved since similar searches of
the previous LIGO run (S4 in 2004)~\cite{S4burstAllSky} and even since the
first year of S5 (S5y1 in 2005--2006)~\cite{S5y1Burst, S5y1BurstHF}. 
This is reflected in an improved strain sensitivity with
$h_{\text{rss}}^{50\%}$ as low (good) as $5.6\times10^{-22}$~Hz$^{-1/2}$ for
certain waveforms (see Table~\ref{table:SGQ9}), compared to best values of
$1.3\times10^{-21}$~Hz$^{-1/2}$ and $6.0\times10^{-22}$~Hz$^{-1/2}$ for S4 and
S5y1 respectively.
The new searches also cover an extended frequency band of 50--6000~Hz.

No plausible gravitational wave candidates have been identified in the
S5y2/VSR1 searches.
Combined with the S5y1 results, which had comparable observation time, this
yields an improved upper limit on the rate of bursts (with amplitudes a few
times larger than $h_{\text{rss}}^{50\%}$) of 2.0 events per year at 90\%
confidence for the 64--2048~Hz band,
and 2.2 events per year for higher-frequency bursts up to 6~kHz.
Thus the full S5/VSR1 upper limit is better than the S5y1 upper limits of 3.75
per year (64--2000~Hz) and 5.4 per year (1--6~kHz), and is more than an order
of magnitude better than the upper limit from S4 of 55 events per year.

We note that the IGEC network of resonant bar detectors set a slightly
more stringent rate limit, $1.5$ events per year at 95\% confidence
level~\cite{astone2003}. However, those detectors were sensitive only
around their resonant frequencies, near 900 Hz, and achieved that rate
limit only for signal amplitudes (in $h_{\text{rss}}$ units) of a few
times $10^{-19}\,\text{Hz}^{-1/2}$ or greater, depending on the signal
waveform.  (See Sec.~X of~\cite{s2burst} for a discussion of this
comparison.)  Further IGEC observations during 6 months of
2005~\cite{IGEC2-2007} improved the rate limit to $\simeq$8.4 per year
for bursts as weak as a few times $10^{-20}\,\text{Hz}^{-1/2}$ but did
not change the more stringent rate limit for stronger bursts.  The
current LIGO-Virgo burst search is sensitive to bursts with $h_\mathrm{rss}$ one to two
orders of magnitude weaker than those which were accessible to the
IGEC detectors.

To characterize the astrophysical sensitivity achieved by the S5y2/VSR1 search,
we calculate the amount of mass, converted into GW burst energy at a
given distance $r_0$, that would be sufficient to be detected by the search
with 50\% efficiency ($M_{\text{GW}}$).
Inverting Eq.~(\ref{eq:hrVmass}), we obtain a rough estimate assuming an
average source inclination angle (i.e.\ $h_\mathrm{rss}^2$ is averaged
over $\cos\iota$):
\begin{equation}\label{eq:SGenergy}
M_{\text{GW}}= \frac{\pi^2 c}{G} \, r_0^2 \, f_0^2 \, h_{\text{rss}}^2 \, .
\end{equation}
For example, consider a sine-Gaussian signal with $f_0=153$~Hz and $Q=9$,
which (from Table~\ref{table:SGQ9}) has
$h^{50\%}_{\text{rss}} = 6.0 \times 10^{-22} \,\text{Hz}^{-1/2}$
for the four-detector network.
Assuming a typical Galactic distance of 10\,kpc,
that $h_{\text{rss}}$ corresponds
to $M_{\text{GW}}=1.8 \times 10^{-8}\,M_\odot$.
For a source in the Virgo galaxy cluster, approximately 16\,Mpc away, the 
same $h^{50\%}_{\text{rss}}$ would be produced by a mass conversion of roughly 
$0.046\,M_\odot$.  These figures are slightly better than for the S5y1
search and a factor of $\sim$5 better than the S4 search.

We also estimate in a similar manner a detection range for GW signals from core-collapse
supernovae and from neutron star collapse to a black hole.  Such
signals are expected to be produced at a much higher frequency (up to
a few kHz) and also with a relatively small GW energy output
($10^{-9}-10^{-5}\,M_\odot{c^2}$). For a possible supernova scenario,
we consider a numerical simulation of core collapse by Ott et
al.~\cite{ott06}.  For the model s25WW, which undergoes an
acoustically driven explosion, as much as $8 \times 10^{-5}\,M_\odot$
may be converted to gravitational waves.  The frequency content
produced by this particular model peaks around $\sim 940$\,Hz and the
duration is of order one second.  Taking this to be similar to a high-$Q$
sine-Gaussian or a long-duration white noise burst, from our detection
efficiency studies we estimate $h_\mathrm{rss}^\mathrm{50\%}$ of 17--22$\times10^{-22}$~Hz$^{-1/2}$, i.e.\ that such a signal could be detected out
to a distance of around $30$~kpc.
The axisymmetric neutron star collapse signals D1 and D4 of Baiotti {\it
  et al.}~\cite{SNwave} have detection ranges (at 50\% confidence) of only
about 25~pc and 150~pc (see Fig.~\ref{fig:D1D4range}, due mainly to their lower
energy ($M_\mathrm{GW} < 10^{-8}~M_\odot$) and also to emitting most of that
energy at 2--6~kHz, where the detector noise is greater.

The Advanced  LIGO and Virgo detectors, currently under 
construction, will increase the detection range of the searches by an order of
magnitude, therefore increasing by $\sim$1000 the monitored volume of the universe. 
With that sensitivity, GW signals from binary mergers are expected to
be detected regularly, and other plausible sources may also be
explored.
Searches for GW burst signals, capable of detecting unknown signal
waveforms as well as known ones, will continue to play a central role
as we increase our understanding of the universe using gravitational
waves.

\begin{acknowledgments}\label{sec:acknowledgements}

The authors gratefully acknowledge the support of the United States National 
Science Foundation for the construction and operation of the LIGO Laboratory, 
the Science and Technology Facilities Council of the United Kingdom, the 
Max-Planck-Society and the State of Niedersachsen/Germany for support of the 
construction and operation of the GEO\,600 detector, 
and the Italian Istituto Nazionale di Fisica Nucleare and the French Centre 
National de la Recherche Scientifique for the construction and operation of 
the Virgo detector. The authors also gratefully acknowledge the support of 
the research by these agencies and by the Australian Research Council, the 
Council of Scientific and Industrial Research of India, the Istituto 
Nazionale di Fisica Nucleare of Italy, the Spanish Ministerio de Educaci\'on 
y Ciencia, the Conselleria d'Economia Hisenda i Innovaci\'o of the Govern de 
les Illes Balears, the Foundation for Fundamental Research on Matter supported 
by the Netherlands Organisation for Scientific Research,
the Polish Ministry of Science and Higher Education, the FOCUS Programme
of Foundation for Polish Science, the Royal Society, 
the Scottish Funding Council, the Scottish Universities Physics Alliance, 
the National Aeronautics and Space Administration, the Carnegie Trust, 
the Leverhulme Trust, the David and Lucile Packard Foundation, the Research 
Corporation, and the Alfred P. Sloan Foundation. 
This document has been assigned LIGO Laboratory document number \ligodoc.

\end{acknowledgments}
\appendix

\section{Data Quality Flags}\label{sec:dataquality}

The removal of poor-quality LIGO data uses the data quality flag (DQF) 
strategy described in the first year analysis~\cite{S5y1Burst}. 
For the second year there are several new DQFs.
New category 2 flags mark high currents in the end test-mass side coils, 
discontinuous output from a tidal compensation feed-forward system,
periods when an optical table was insufficiently isolated from ground noise, 
and power fluctuations in lasers used to thermally control the
radius of curvature of the input test masses. 
A flag for overflows of several of the main photodiode readout sensors that was 
used as a category 3 flag in the first year was promoted to category 2.
New category 3 flags mark noise transients from light scattered from 
H1 into H2 and vice versa, large low-frequency seismic motions,  
the optical table isolation problem noted above, 
periods when the roll mode of an interferometer optic was excited,
problems with an optical level used for mirror alignment control,
and one period when H2 was operating with degraded sensitivity. 
The total ``dead time'' (fraction of live time removed)
during the second year of S5 due to category 1 DQFs was
2.4\%, 1.4\%, and $<$0.1\% for H1, H2, and L1, respectively. Category 2 DQF 
dead time was 
0.1\%, 0.1\%, and 0.6\%, and category 3 DQF dead time was 4.5\%, 5.5\%, and 7.7\%.
Category 4 flags, used only as additional information for follow-ups of
candidate events (if any), typically 
flag one-time events identified by Collaboration members on duty in the
observatory control rooms, and thus are 
quite different between the first and second years. 

Virgo DQFs are defined by study of the general behavior of the detector, 
daily reports from the control room, online calibration information, and 
the study of loud transient events generated online from the uncalibrated 
Virgo GW channel by the Qonline \cite{Blackburn08} program.
Virgo DQFs include out-of-science mode, hardware injection periods, and
saturation of the current flowing in the coil drivers.
Most of them concern a well identified detector or data acquisition problem, 
such as the laser frequency stabilization process being off, photodiode 
saturation, calibration line dropouts, and loss of synchronization of the 
longitudinal and angular control.  Some loud glitches and periods of
higher glitch rate are found to be due to environmental conditions, such 
as increased seismic noise (wind, sea, and earthquakes), and 50 Hz power line 
ground glitches seen simultaneously in many magnetic probes.
In addition, a faulty piezo-electric driver used by the beam monitoring 
system generated glitches between 100 and 300 Hz, and a piezo 
controlling a mirror on a suspended bench whose cabling was not well 
matched caused glitches between 100 and 300 Hz and between 600 and 700 Hz.
The total dead time in VSR1 due to category 1 DQFs was 1.4\%. Category 2 DQF 
dead time was 2.6\%, and category 3 DQF dead time was 2.5\% \cite{Leroy09}.

\section{Event-by-event vetoes}\label{sec:vetoes}

Event-by-event vetoes discard gravitational-wave channel noise
events using information from the many environmental and interferometric
auxiliary channels which measure non-GW degrees of freedom. Our procedure for
identifying vetoes in \sfyt~and VSR1 follows that used in
\sfyo~\cite{S5y1Burst}. 
Both the GW channels and a large number of auxiliary channels are
processed by the KleineWelle (KW)~\cite{blackburn2005} algorithm, which 
looks for excess power transients.  Events from the auxiliary channels 
which have a significant statistical correlation with the events in the 
corresponding GW channel are used to generate the veto time intervals.
Candidate events identified by the search algorithms are rejected if they 
fall inside the veto time intervals.

Veto conditions belong to one of two categories which follow the same notation
used for data quality flags. Category 2 vetoes are a conservative set of vetoes
targeting known electromagnetic and seismic disturbances at the LIGO and Virgo
sites. These are identified by requiring a coincident observation of an
environmental disturbance across several channels at a particular site. The
resulting category 2 data selection cuts are applied to all analyses described
in this paper, and remove $\sim$0.2\% of analyzable coincident live time.
Category 3 vetoes make use of all available auxiliary channels shown not to
respond to gravitational waves. An iterative tuning method is used to 
maximize the number of vetoed noise events in the gravitational-wave
channel while removing a minimal amount of time from the analysis.  
The final veto list is applied to all analyses
below 2048 Hz, removing $\sim$2\% of total analyzable coincident live time.

An additional category 3 veto condition is applied to Virgo triggers, 
based on the ratio of the amplitude of an event as measured in the 
in-phase (P) and quadrature (Q) dark port demodulated signals.  
Since the Q channel should be insensitive to a GW signal, large Q/P ratio 
events are vetoed.  This veto has been verified to be safe using 
hardware signal injections \cite{Ballinger09}, with a loss of live time 
of only 0.036\%.

\section{EGC burst search}
\label{sec:EGC}

The Exponential Gaussian Correlator (EGC) pipeline is based on a matched 
filter using exponential Gaussian templates \cite{clapson08, acernese09a}, 
\begin{equation}
\Phi(t)=\exp\left(- \frac{t^2}{2\tau_0^2} \right) e^{2 \pi i f_0 t} \, ,
\end{equation}
where $f_0$ is the central frequency and $\tau_0$ is the duration.
Assuming that real GWBs are similar to sine-Gaussians, EGC 
cross-correlates the data with the templates,
\begin{equation}
C(t) = \frac{1}{N} \int_{-\infty}^{+\infty} \frac{\tilde{x}(f) \tilde{\Phi}^*(f)}{S(f)} e^{2\pi i ft} df \, .
\end{equation}
Here $\tilde{x}(f)$ and $\tilde{\Phi}(f)$ are the Fourier transforms 
of the data and template, and $S(f)$ is the two-sided noise power 
spectral density.  $N$ is a template normalization factor, defined as 
\begin{equation}
N=\sqrt{\int_{-\infty}^{+\infty} \frac{|\Phi(f)|^2}{S(f)} df} \, .
\end{equation}
We tile the parameter space $(f_0, Q_0\equiv2\pi \tau_0 f_0)$ using 
the algorithm of \cite{arnaud03}.  The minimal match is 72\%, while 
the average match between templates is 96\%.  The analysis covers 
frequencies from 300 Hz to 5 kHz, where LIGO and Virgo have comparable 
sensitivity.  $Q_0$ varies from 2 to 100, covering a large range of 
GW burst durations. 

The quantity $\rho=\sqrt{2 |C|^2}$ is the signal-to-noise ratio 
(SNR), which we use to characterize the strength of triggers in 
the individual detectors.  The analysis is performed on times 
when at least three of the four detectors were operating.
Triggers are generated for each of the four detectors and kept 
if the SNR is above 5.  In order to reduce the background, 
category 2 DQFs and vetoes are applied, followed by several other tests.  
First, triggers must be coincident in both time and frequency 
between a pair of detectors.  The time coincidence window is the 
light travel time between the interferometers plus a conservative 
10 ms allowance for the EGC timing accuracy.  The frequency coincidence
window is selected to be 350~Hz. 
Second, events seen in coincidence in H1 and H2 with a unexpected  
ratio in SNR are discarded (the SNR in H1 should be approximately 
2 times that in H2).
Surviving coincident triggers are ranked according to the combined 
SNR, defined as 
\begin{equation}
\rho_{comb}=\sqrt{\rho_1^2 + \rho_2^2} \, ,
\end{equation}
where $\rho_1$ and $\rho_2$ are the SNR in the two detectors.  Third, 
a threshold is applied on $\rho_1$ and $\rho_2$ to reduce the trigger 
rate in the noisier detector.  This lowers the probability that a 
detector with a large number of triggers will generate many 
coincidences with a few loud glitches in the other detector.  Finally, 
for each coincident trigger we compute the SNR disbalance measure 
\begin{equation}
\Lambda = \frac{\rho_{comb}}{\rho_{comb}+|\rho_1 - \rho_2|} \, .
\end{equation}
This variable is useful in rejecting glitches in a pair of co-aligned 
detectors with similar sensitivity, and so is used primarily for pairs 
of triggers from the LIGO detectors.

The background is estimated for each detector pair by time shifting  
the trigger lists. 
200 time slides are done for H1H2L1V1, and more for the three-detector 
networks due to their shorter observation times (see Table \ref{tab:egc}).
The thresholds applied to $\rho_1$, $\rho_2$ and $\Lambda$ are tuned 
for each detector pair to maximize the average detection efficiency 
for sine-Gaussian waveforms at a given false alarm rate.
Once the $\rho_1$, $\rho_2$ and $\Lambda$ thresholds are applied, all 
trigger pairs from the network are considered together and $\rho_{comb}$ 
is used as the final statistic to rank the triggers.  
A threshold is placed on $\rho_{comb}$, 
chosen to give a low false alarm rate.  More precisely, as we observe 
an excess of loud glitches with $f_0<400$ Hz, we use different 
thresholds depending on the frequency of the coincident triggers. 
Below 400 Hz, the false alarm rate is tuned to 1 event per 10 years. 
Above, the threshold for each network is set to give a maximum of 
0.05 events expected from background for that network.  An exception  
is made for H1L1V1, where the maximum number is chosen to be 0.01 
events because of its shorter observation time.  The final thresholds for 
each network are given in Table \ref{tab:egc}.

\begingroup
\squeezetable
\begin{table}
\begin{tabular}{|c|c|c|c|c|}
\hline
Network   & Obs.\ time  & \# lags & FAR                              & $\rho_{comb}$ \\ 
 	  &	[days]	   &		& 				&		\\	
\hline
H1H2L1V1  & 66.6            & 200     & $<$ 400 Hz: 1 event in 10 years  & 69.8 \\
	  &                 & 	      & $>$ 400 Hz: 0.05 events          & 21.0 \\
H1H2L1    & 18.3            & 1000    & $<$ 400 Hz: 1 event in 10 years  & 80.9 \\
          &                 &         & $>$ 400 Hz: 0.05 events          & 10.0 \\
H1H2V1    & 15.9            & 1000    & $<$ 400 Hz: 1 event in 10 years  & 89.6 \\
	  &		    &	      & $>$ 400 Hz: 0.05 events          & 15.4 \\
H1L1V1    & 4.5             & 2000    & $<$ 400 Hz: 1 event in 10 years  & 67.9 \\
	  &		    &	      & $>$ 400 Hz: 0.01 events          & 24.2 \\
\hline
\end{tabular}
\caption{\label{tab:egc} Thresholds and background tuning information for all the networks studied by the EGC pipeline.}

\end{table}
\endgroup

\section{$\Omega$-Pipeline burst search}
\label{sec:omega}

The $\Omega$-Pipeline is essentially identical to QPipeline, which was 
used in previous LIGO S5 searches \cite{S5y1Burst, S5y1BurstHF}.  
QPipeline has since been integrated into a larger software suite, 
with a change in nomenclature but no significant change in 
methodology.  Since this approach is discussed in detail in \cite{S5y1Burst,chatterjiThesis}, we provide only a summary here.

The $\Omega$-Pipeline, like EGC, functions as a matched-filter search on a 
single-interferometer basis.  The data stream is whitened by linear 
predictive filtering \cite{linearPrediction}, then projected onto a 
template bank of complex exponentials.  These 
templates are similar to those used by EGC, parametrized by central 
time $\tau_0$, central frequency $f_0$, and quality factor $Q_0$, but 
use bisquare windows rather than Gaussian windows.
The template spacing is also different, selected for computational speed, 
rather than for strict mathematical optimization as in EGC.  
The $\Omega$ template bank has a minimal match of 80\%, and covers a frequency 
range from 48 Hz to 2048 Hz and a $Q$ range from 2.35 to 100. 

The significance of a single-interferometer trigger is given by its 
normalized energy $Z$, defined as the ratio of the squared magnitude 
of $X$ (the projection onto the best-matched template) for that trigger 
to the mean-squared magnitude of other templates with the same $f_0$ and 
$Q_0$. For Gaussian white noise, $Z$ is exponentially distributed and 
related to the matched filter SNR $\rho$ by
\begin{equation}\label{eqn:Z}
Z = |X|^2/\langle|X|^2\rangle = \rho^2/2 \, .
\end{equation}
$Z$ is used to rank L1 and V1 triggers.

For H1 and H2, $\Omega$-Pipeline takes advantage of their co-located 
nature to form two linear combinations of the data streams.  
The first of these, the {\em coherent} stream H$_+$, is the sum of the strains in the two
interferometers weighted by their noise power spectral densities.
We define the coherent energy $\Zcoh$ following 
(\ref{eqn:Z}).  We also define the correlated energy $\Zcor$, 
which is obtained by removing the contribution to $\Zcoh$ 
from H1 and H2 individually and leaving only the cross-correlation 
term\footnote{This statement is actually only approximately correct due to 
complications related to normalization.  See \cite{S5y1Burst} for more details.}.
The H1H2 cuts are based on $\Zcor$, because 
it is less susceptible than $\Zcoh$ to instrumental glitches, 
so providing better separation between signal and noise.
The second stream, the \emph{null} stream H$_-$, is the difference 
between the strains in H1 and H2.  The normalized energy
$Z_{H_-}$ should be small for a gravitational wave, but generally 
much larger for an instrumental glitch.  We therefore veto coherent 
stream triggers which are coincident in time and frequency with 
null stream triggers.

We require triggers to be coincident in at least two detectors.  
The interferometer combinations analyzed are shown in 
Table~\ref{tab:omegacuts}. (Note that because of the coherent 
analysis of H1 and H2, both must be operating for data from 
either to be analyzed.)  Triggers are required to be coincident 
in both time and frequency as follows:
\begin{eqnarray}
|T_1 - T_2| & < & T_c + \frac12\,\mathrm{max}(\sigma_1 , \sigma_2) \\
|F_1 - F_2| & < & \frac12\,\mathrm{max}(b_1 , b_2) \, .
\end{eqnarray}
Here $T$ and $F$ are the central time and frequency of the triggers, 
$\sigma$ and $b$ are their duration and bandwidth, and   
$T_{c}$ is the light travel time between the interferometers.  

The background for each detector pair is determined by time-shifting 
the triggers from one detector.  We use 1000 shifts for each pair, 
except H1-H2.
Only 10 shifts between H1 and H2 are used because the coherent 
analysis requires each shift to be processed independently, 
substantially increasing the computational cost.  Also, time 
shifts between H1 and H2 are less reliable because they miss 
correlated background noise from the common environment.
For all pairs, triggers below and above 200 Hz are treated separately
because of the different characteristics of the glitch 
populations at these frequencies.
  
\begin{figure}[!h]
\begin{center}
\mbox{
\includegraphics[width=0.5\textwidth]{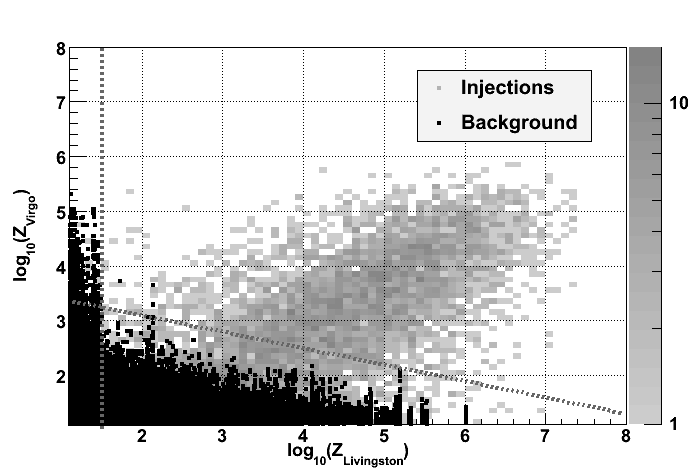}}
\caption{\label{fig:omegadistribution}
Distribution of background and injection triggers below 200 Hz 
after category 3 DQFs and vetoes for L1-V1 pair.  The dashed 
lines show the final normalized energy thresholds.}
\end{center}
\end{figure}

Normalized energy thresholds are set separately for each detector 
combination and frequency range such that there is less than a 5\% 
probability of a false alarm  
after category 3 DQFs and vetoes. Table \ref{tab:omegacuts} shows
the thresholds and surviving events in timeslides for each  
combination.  Fig.~\ref{fig:omegadistribution} shows background 
and injection triggers and the energy thresholds for one 
interferometer pair. 

{\small{ 
\begin{table}[htb]
\begin{tabular}{|c|c|c|}
\hline
Detector combination & threshold &events in 1000\\
and frequency band & &timeslides \\ 
\hline
H1H2L1 $<$ 200 Hz & $\Zcor>37$, $Z_{L1}>13$               & 14 \\
H1H2L1 $>$ 200 Hz & $\Zcor>13$, $Z_{L1}>13$               & 16 \\
H1H2V1 $<$ 200 Hz & $\Zcor>22$, $Z_{V1}>13$               & 9 \\
H1H2V1 $>$ 200 Hz & $\Zcor>14$, $Z_{V1}>13$               & 0 \\
L1V1   $<$ 200 Hz & $Z_{L1}>32$ and $Z_{V1}>$             & 4 \\
                  & $(4.7\times 10^{-13} Z_{L1})^{-0.3}$  & \\
L1V1   $>$ 200 Hz & $Z_{L1}>30$ and $Z_{V1}>$             & 5 \\
                  & $(6.9\times 10^{-12} Z_{L1})^{-0.27}$ & \\
H1H2   $<$ 200 Hz & $\Zcor>80$                            & 0 (10 slides)\\
H1H2   $>$ 200 Hz & $\Zcor>30$                            & 0 (10 slides)\\
\hline
total events      &                                       & 48\\
\hline
\end{tabular}
\caption{\label{tab:omegacuts} Thresholds on normalized energy for the various detector combinations.}
\end{table}
}}

\section{Coherent WaveBurst Search}
\label{sec:cWB}

Coherent WaveBurst (cWB) is a coherent algorithm for detecting 
gravitational-wave bursts.  It constructs a least-squares fit 
of the two GW polarizations to the data from the different detectors 
using the constraint likelihood method \cite{klimenko05}.  
The cWB algorithm was first used
in search for gravitational wave bursts in the LIGO-GEO network \cite{LIGO-GEO-S4}.
More recently it has been used in the LIGO S5 first-year low-frequency search \cite{S5y1Burst}, 
and detailed descriptions of the algorithm can be found there and in 
\cite{klimenko05,klimenko08}. 

The cWB analysis in this search covers frequencies from 64 Hz 
to 6.0 kHz, with the data processing split into two bands.  The 
low-frequency (LF) band (64 Hz to 2.0 kHz) contains the most sensitive 
(but also the most non-stationary) data.  The high-frequency (HF) band 
(1.28 kHz to 6.0 kHz) is dominated by the shot noise of the 
detectors and is much less polluted by environmental and 
instrumental transients.  Splitting the analysis into two bands 
is convenient for addressing the different noise characteristics 
in these bands.  It also eases the computational cost.  
The overlap of the bands is used to cross-check the results and 
to preserve the sensitivity to wide-band signals near the
boundary between the bands.

The cWB analysis is performed in several steps.  First, the data 
are decomposed into Meyer wavelets.  Time-frequency resolutions of
($ 8\times1/16$, $16\times1/32$, $32\times1/64$, $64\times1/128$, 
$128\times1/256$, $256\times1/512$ [Hz $\times$ s]) are used for 
the low-frequency search and ($12.5\times1/25$, $25\times1/50$, 
$50\times1/100$, $100\times1/200$, $200\times1/400$, $400\times1/800$  
[Hz $\times$ s]) for the high-frequency search.  The data are processed 
with a linear predictor error filter to remove power lines, violin modes
and other predictable data components.  Triggers are identified as 
sets of wavelet pixels among the detectors containing excess power 
at time delays consistent with a gravitational wave from a physical 
sky position.  For each trigger, trial incoming sky locations are 
sampled with $1^\circ$ resolution, and various coherent 
statistics are computed.  These include the maximum likelihood ratio 
$L_m$  (a measure of the sum-squared matched-filter SNR detected in the
network), the network correlated amplitude $\eta$, the network correlation coefficient (${cc}$),
the energy disbalance statistics 
$\Lambda$, $\Lambda_{\text{HH}}$ and the penalty factor $P_f$.
(Each of these statistics is described in detail in \cite{S5y1Burst}.) 
The trial sky position giving the largest $cc \cdot L_m$ is 
selected as the best-guess incident direction, and the coherent 
statistics for this position are recorded.  Finally, several 
post-production selection cuts are applied to the triggers to reduce 
the background. 

\begin{figure}
\begin{center}
\mbox{
\includegraphics*[width=0.5\textwidth]{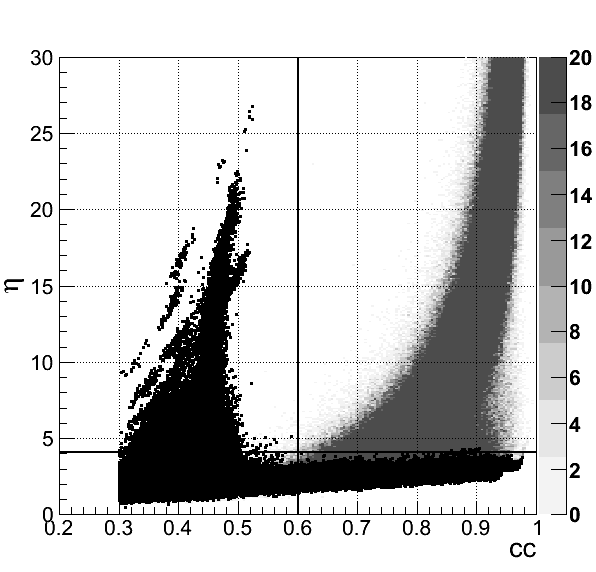}}
\caption{
Distribution of background triggers (black dots) after category 2 
DQFs and vetoes for the L1H1H2 network for the high-frequency search, 
with $Q=9$ sine-Gaussians injections (gray dots).  The dashed lines 
show the thresholds on $\eta$ and $cc$ chosen for this network. 
}
\label{fig:cwb_cc_rho}
\end{center}
\end{figure}

Two groups of selection cuts are used in cWB. First, cuts on ${cc}$, 
$\Lambda$, $\Lambda_{\text{HH}}$ and $P_f$ are used to distinguish 
noise outliers from genuine GW signals. The most powerful consistency
cut is based on the network correlation coefficient $cc$.  For example, 
Figure~\ref{fig:cwb_cc_rho} shows a scatter plot of background triggers 
as a function of $\eta$ and $cc$.  Strong outliers (with large values 
of $\eta$) are characterized by low values of $cc$ and are 
well separated from simulated signals.  Additional selection cuts are 
based on the energy disbalance statistics $\Lambda_{\text{NET}}$, 
$\Lambda_{\text{HH}}$, and $P_f$. They are used to reject specific 
types of background events, such as H1-H2 correlated transients. 

The second, final cut is on the network correlated amplitude $\eta$, 
which characterizes the significance of the triggers.
Due to different characteristics of the background noise during the run and 
in the different frequency bands, the threshold on $\eta$ is selected 
separately for each network configuration and frequency band to give 
false alarm probabilities of a few percent.  In particular, in the 
low-frequency search separate $\eta$ thresholds are used for triggers 
below and above 200 Hz.  Table~\ref{tab:cuts} shows the thresholds used 
in the analysis. 

\begin{table}[htbp]  	
\begin{center}
\begin{tabular}{|c|c c c c c c|c c c|}
\hline
 & \multicolumn{6}{|c|}{\textbf{cWB LF}  } & \multicolumn{3}{c|}{\textbf{cWB HF} } \\ 

 & $cc$ &  \textbf{$\eta_{1}$}  & \textbf{$\eta_{2}$}  & \textbf{$\Lambda_\text{HH}$}  & \textbf{$\Lambda_\text{NET}$} & \textbf{$P_f$} & $cc$ & \textbf{$\eta$} & \textbf{$P_f$}\\ 
\hline
\hline
\textbf{H1H2L1V1} & 0.5    & 4.5  & 4.1  & 0.3 & 0.35 & 0.6 & 0.6& 4.3 & 0.6\\ 
\hline
\textbf{H1H2L1}   &  0.6 & 6.0 & 4.2  & 0.3 & 0.35 & 0.6 & 0.6 & 4.1 & 0.6\\ 
\textbf{H1H2V1}   &   -  & -  & -  & - & - & - & 0.7 & 4.6 & 0.6\\ 
\textbf{H1L1V1}   &  0.5   & 5.0   &  5.0 & - & 0.7  & 0.6 & 0.6 & 6.0 & 0.6\\ 
\hline
\textbf{H1H2}     &  0.6 & 6.0  & 4.1 & 0.3 & 0.35 & 0.6 & 0.5 & 5.0 & 0.6 \\ 
\textbf{H1L1}     &  0.6 & 9.0  & 5.5 & - & 0.35 & 0.6 & - & - & - \\ 
\textbf{H2L1}     &  0.6 & 6.5  & 5.5 & - & 0.35 & 0.6 & - & - & - \\ 
\hline
\end{tabular}
\end{center}
\caption{
Thresholds for each network for the cWB low- and high-frequency searches. 
Different thresholds on $\eta$ are used for triggers below 200 Hz 
($\eta_1$) and above 200 Hz ($\eta_2$) due to the different characteristics 
of the LIGO background noise in these 
frequency ranges.  No energy disbalance cuts ($\Lambda_\text{NET}$, 
$\Lambda_\text{HH}$) are applied in the high-frequency search. 
In addition, a penalty factor cut of $P_f>0.6$ is applied to all 
network configurations and searches. 
}  
\label{tab:cuts} 
\end{table}

The background is estimated separately on each segment of data processed.  
The cWB algorithm forms circular data buffers and shifts one detector with respect to 
the others, repeating the analysis hundreds of times on the time-shifted 
data. 
Table \ref{tab:lags} shows the number of lags and accumulated 
background observation time for the various cWB searches and network 
configurations.  The background data sets are used for tuning 
of the cWB selection cuts and also for estimation of the 
significance of the foreground events. For example, to estimate 
the significance of the blind injection identified by cWB, we 
generated a background sample with observation time equivalent to 
approximately 1000 H1H2L1 \sfyt/VSR1 data sets. 

\begingroup
\squeezetable
 \begin{table}[htbp]

\begin{center}
\begin{tabular}{|c|c|c|c|c|c|c|}
\hline
 & \multicolumn{3}{|c|}{\textbf{cWB LF Background}  } & \multicolumn{3}{c|}{\textbf{cWB HF Background} } \\ 
 &  \# lags& Obs time& FAR & \# lags& Obs time & FAR\\ 
  & & [$year$]& [$year^{-1}$] & & [$year$]  & [$year^{-1}$]\\
\hline
\textbf{H1H2L1V1} & 200   & 34.8 & 0.3 & 96  & 17.6 & 0.17\\ 
\hline
\textbf{H1H2L1}   & 1000  & 499.9 & 0.1 & 96 & 31.7 & 0.09\\ 
\textbf{H1H2V1}   & -    & -  & - & 96  & 3.7 & 0.27\\ 
\textbf{H1L1V1}   & 200    & 1.8   & 3.3 & 288  & 3.0 & 0.33  \\ 
\hline
\textbf{H1H2}     & 200   & 28.2    & 0.04 & 192  & 17.2 & 0.06   \\ 
\textbf{H1L1}     & 200   & 4.6     & 2.0 & - & -  & - \\ 
\textbf{H2L1}     & 200   & 1.6     & 0.6 & - & -  & - \\ 
\hline
\end{tabular}
\end{center}
\caption{\label{tab:lags}
Background observation time and false alarm rates for each network for the 
cWB low- and high-frequency searches. 
}  
\end{table}
\endgroup

\bibliographystyle{apsrev}

\end{document}